\documentclass[aip,jap,reprint]{revtex4-1}

\usepackage{amsmath}
\usepackage[utf8]{inputenc}
\usepackage{academicons}
\usepackage{xcolor}
\usepackage{graphics}
\usepackage{graphicx}
\usepackage[T1]{fontenc} 
\usepackage[rightcaption]{sidecap}
\sidecaptionvpos{figure}{c}

\usepackage[breaklinks=true]{hyperref}
\usepackage{tikz}
\usepackage[all]{hypcap}
\usepackage{orcidlink}
\definecolor{JAPblue}{rgb}{0, 0.682, 0.937}
\hypersetup{
	colorlinks = true,
	linkcolor = {JAPblue},
	citecolor = {JAPblue},
	urlcolor = {JAPblue},
}
\usepackage{amsfonts}

\usepackage{upgreek}
\usepackage{amsmath}
\usepackage{mathtools}

\begin{document}

\title{Ultrafast temperature diagnosis of dynamically compressed matter using millielectronvolt inelastic x-ray scattering beyond the first Brillouin zone}

\author{P.G. Heighway~\orcidlink{0000-0001-6221-0650}}\email{patrick.heighway@physics.ox.ac.uk}
\affiliation{Department of Physics, Clarendon Laboratory, University of Oxford, Parks Road, Oxford OX1 3PU, UK\looseness=-1}

\author{J.S. Wark~\orcidlink{0000-0003-3055-3223}}
\affiliation{Department of Physics, Clarendon Laboratory, University of Oxford, Parks Road, Oxford OX1 3PU, UK\looseness=-1}

\date{\today}

\begin{abstract}
We present calculations of the millielectronvolt-scale x-ray scattering spectra of multilayered dynamic-compression targets comprising an unstructured ablator layer and a crystalline, textured sample layer. Our model builds on the classic formulation of x-ray thermal diffuse scattering by Warren [B.\,E.\,Warren, Acta Crystallogr.\,\textbf{6}, 803 (1953)] and includes both elastic and first-order (single-phonon) inelastic scattering contributions to the dynamic structure factor $S(\mathbf{q},\omega)$. We focus on the umklapp scattering regime (i.e., at momentum transfers outside the first Brillouin zone) where the ablator scattering that threatens to overwhelm the inelastic scattering from the crystalline layer of interest is suppressed. We show that, despite the considerably more complex structure of the inelastic scattering spectra in this intermediate-$q$ regime, it is still possible to reliably deduce the temperature of the crystal using Dornheim's Laplace-transform--based formalism [Dornheim \textit{et al.}, Phys.\,Plasmas \textbf{30}, 042707 (2023)], regardless of the details of the sample's texture.
\end{abstract}

\maketitle

\section{Introduction}

\begin{SCfigure*}
    \centering
    \includegraphics{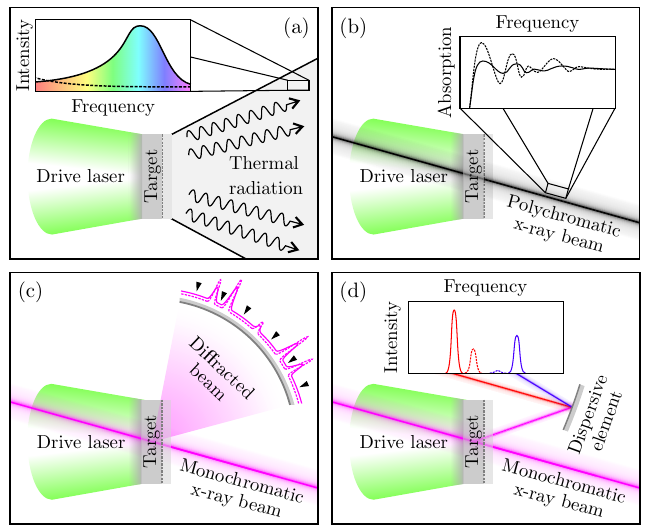}
    \caption{Main temperature diagnostics used in dynamic-compression science. (a) Optical pyrometry, whereby temperature is calculated from a target's rear-surface thermal emission spectrum. (b) Extended x-ray absorption fine structure (EXAFS), in which temperature is inferred from the depth of the modulations of the target's x-ray absorption coefficient marginally above an absorption edge. (c) Thermal diffuse scattering (TDS), whereby temperature is derived from the intensity of the target's diffraction pattern between the Bragg peaks, where inelastic scattering from phonons dominates. (d) Millielectronvolt inelastic x-ray scattering (meV-IXS), in which the TDS at a certain momentum transfer is spectrally dispersed, and the temperature is calculated from the relative probabilities of phonon emission or absorption.}
    \label{fig:thermometers}
\end{SCfigure*}

Modern dynamic-compression techniques bring planetary interior conditions within easy experimental reach. New diode-pumped, gigawatt-power, long-pulse lasers\cite{Mason2018} can ablatively drive solid specimens to multimillion-atmosphere pressures at a repetition rate now approaching the hertz level\cite{Gorman2024}. Ultrabright light sources like x-ray free-electron lasers (XFELs) allow us to probe these short-lived, extreme thermodynamic states \textit{in situ}, and thus capture compression-induced microphysical processes that reshape atomistic structure on subpicosecond timescales. Extreme phenomena that have been elucidated by this uniquely powerful pump-probe configuration include: plastic deformation\cite{Suggit2012,Milathianaki2013,Wehrenberg2017,Sharma2020}, dislocation dynamics\cite{Avraam2023}, and damage\cite{Coakley2020} at meteoric-impact--like strain rates; pressure-induced solid-solid phase transitions in a variety of metals\cite{Kalantar2005,Sharma2020b,Singh2024,Heighway2024}, nonmetals\cite{McBride2019,Tracy2019}, and minerals\cite{Tracy2018,Wicks2024}; shock-induced melting\cite{Gorman2015,Singh2023} and pressure-dependent liquid structure at terrestrial core conditions\cite{Crepisson2025b}; and subnanosecond amorphization of crystal structure\cite{Hernandez2020,Gleason2022,Crepisson2025}. Understanding these phenomena is essential for predicting planetary formation and stratification, modeling hypervelocity collision events in defense and aerospace scenarios, and tailoring compression pathways to access exotic, functional phases of matter with extraordinary mechanical\cite{NguyenCong2024} and electrical\cite{Sun2021} properties.

While density is readily derived from x-ray diffraction-based measurements of its lattice constants, and pressure from time-resolved velocimetric measurements of its rear-surface motion, the \emph{temperature} of dynamically compressed solid matter is notoriously difficult to determine. The extremely short timescale of laser-based dynamic-compression experiments $(<\text{ns})$ results in near-adiabatic thermodynamic conditions that render conventional thermal-conduction--based thermometers impracticable. Ultrafast thermometric techniques are therefore based on radiation (see Fig.~\ref{fig:thermometers}). The oldest and best-developed such technique, optical pyrometry\cite{Hereil2000,Swift2007,Millot2018} [Fig.~\ref{fig:thermometers}(a)], is generally workable only at temperatures approaching 10\textsuperscript{4}~K, as the small laser-target interaction volume $(<\text{mm}^3)$ will otherwise emit too few photons to yield a meaningful signal. To meet the grand challenge of performing high-precision, \textit{in-situ} temperature measurements of high-energy-density matter within the strict spatiotemporal limits of a laser-compression experiment, the community has turned to x-ray--based diagnostic techniques that exploit some of the brightest lightsources on Earth.

The most mature x-ray--based temperature diagnostic as applied to dynamically compressed matter is based on measurement of extended x-ray absorption fine structure (EXAFS) [Fig.~\ref{fig:thermometers}(b)]. Here, one illuminates the target with a broadband x-ray source (conventionally either a synchrotron\cite{Turneaure2022} or a laser-driven plasma `backlighter'\cite{Sio2023}) whose spectral range includes an absorption edge, and calculates the absorption coefficient as a function of photon energy marginally above that edge. The wavefunction of each ejected photoelectron interacts with and is partially scattered by the atoms surrounding its host atom, causing modulations in the photon absorption coefficient whose amplitude is sensitive to the degree of spatial correlation between neighboring atoms. This correlation is, in turn, determined by the typical thermal displacement of the atoms from their equilibrium positions, $\langle u^2 \rangle$, thus giving a proxy measurement of their collective temperature $T$. The relationship between $\langle u^2 \rangle$ and $T$ is usually calculated via the Debye model, wherein the steepness of the potential well to which each atom is confined is parametrized by the Debye temperature $\Theta_D$, a density-dependent material property whose value must usually be modeled due to a general paucity of experimental measurements in the high-energy-density regime. Hence, in addition to relatively large statistical errors (typically no lower than 20\%\cite{Turneaure2022,Sio2023}), EXAFS-based temperature measurements are subject to inherent systematic uncertainty arising from the need to specify a Debye temperature.

Belonging to the same class of Debye-temperature-reliant techniques is the relatively new approach of measuring x-ray thermal diffuse scattering (TDS) [Fig.~\ref{fig:thermometers}(c)]. As the temperature of a crystalline sample increases, the intensity of its Bragg peaks (which is to say, the elastic component of its x-ray diffraction pattern) will decrease due to increasing thermal displacement of its atoms. As the elastic scattering diminishes, the inelastic scattering, which manifests as a diffuse `background' signal sitting beneath and between the Bragg peaks, increases at an equal rate. Much like EXAFS, the temperature scaling of the TDS intensity can be predicted if one knows the density-dependent Debye temperature $\Theta_D$. The present authors recently demonstrated the viability of ultrafast TDS-based thermometry in an experiment at the European XFEL (EuXFEL). Here, the authors gathered single-shot, femtosecond x-ray diffraction patterns from polycrystalline copper foils shock-compressed to up to 140~GPa, and, by fitting the thermal diffuse components of the signal to a simple model owed in part to Warren\cite{Warren1953}, extracted Hugoniot temperatures consistent with the predictions of high-pressure thermal equations of state, to within a statistical error comparable with that of EXAFS\cite{Wark2025} and with negligible sensitivity to crystallographic texture\cite{Heighway2025}. While this new technique shows promise, its predictions are ultimately beholden to the accuracy of the modeled Debye temperature.

However, this model dependence can be circumvented entirely if one examines not only the total intensity of the TDS, but its spectral content [Fig.~\ref{fig:thermometers}(d)]. At relatively low momentum transfers (often referred to in the shorthand as `forward scattering'), inelastic x-ray scattering is realized by either the emission or absorption of a single phonon. For a given phonon energy, the relative probability of these two events is controlled exclusively by thermodynamic temperature. Hence, by spectrally dispersing the TDS signal at a fixed momentum transfer and calculating the ratio of the up- and down-scattering peak intensities, the sample's temperature may be derived unambiguously, without appeal to an underlying model of the phonon structure; this technique is referred to as millielectronvolt inelastic x-ray scattering (meV-IXS). The feasibility of fielding this diagnostic at an XFEL has been proven in a small number of pioneering studies\cite{McBride2018,Descamps2020a,Wollenweber2021} on undriven (ambient-pressure) targets. The main impediment to realizing meV-IXS--based temperature measurements of dynamically compressed matter is the exceptionally small scattering cross-section, which, despite the unmatched spectral brightness of an XFEL, requires that spectra from hundreds of identically driven targets be gathered and averaged to yield a single, smooth signal from which a  temperature may be inferred. Such a workflow, though ambitious and experimentally demanding, is now becoming practicable with stable, high-repetition-rate drive lasers like the DiPOLE 100-X system\cite{Mason2018} and rapid-throughput target delivery platforms\cite{Prencipe2017,Smith2022}. The first model-agnostic, femtosecond thermometer for dynamically compressed matter that meV-IXS represents is thus within reach.

Because the technique is still at the proof-of-principle stage, XFEL-based meV-IXS studies undertaken to date have rightly focused on scenarios resulting in the simplest possible spectra. Specifically, they concern single-component, (quasi-)single-crystal targets probed at momentum transfers within the first Brillouin zone. The result is an meV-IXS spectrum comprising up- and down-scattering peaks arising from a single phonon mode only, with negligible background from extraneous scattering sources to obscure them. A `real' experiment is unlikely to be so simple. The sort of dynamic-compression target that can practicably be produced in the volumes required to supply an extremely photon-hungry meV-IXS experiment will likely comprise a mass-producible, polycrystalline layer of the material of interest bonded to a low-cost, low-$Z$ (likely plastic) sacrificial ablator. The reciprocal space of such a target is considerably richer than that of an isolated single crystal. In real terms, this means the inelastic scattering from the sample of interest may have to compete with secondary, potentially much more intense sources of scattering that could easily overwhelm it. We need a deeper, quantitative understanding of the nature of this competition if we are to realize meV-IXS thermometry in a forward-scattering geometry.

The purposes of this paper are threefold. We first take the opportunity in Sec.~\ref{sec:theory} to present a simple, transparent model of single-phonon meV-IXS from cubic polycrystals both with and without crystallographic texture, and benchmark its predictions against Warren's classic model of thermal diffuse scattering\cite{Warren1953} (i.e., spectrally unresolved meV-IXS). In Sec.~\ref{sec:analysis}, we then calculate millielectronvolt-scale scattering spectra from representative multilayered dynamic-compression targets. We quantify the competition between elastic and inelastic scattering at momentum transfers both inside and outside the first Brillouin zone. Using a Laplace-transform--based approach built by Dornheim \textit{et al.}~\cite{Dornheim2022,Dornheim2023}, we show that the temperature of textured polycrystals can be reliably deduced from their meV-IXS spectra, provided the spectra are measured outside the first Brillouin zone where ablator scattering is relatively weak. Finally, we discuss general prospects for using meV-IXS as a temperature diagnostic for dynamically compressed solids in Sec.~\ref{sec:discussion} before concluding in Sec.~\ref{sec:conclusion}.

\section{Millielectronvolt scattering model\label{sec:theory}}

We begin by outlining our model for the meV-IXS spectra generated by a monochromatic x-ray beam scattered by a textured cubic polycrystal. We will consider only the two most important contributions to the scattering at modest temperature and momentum transfer, namely the elastic and the first-order (single-phonon) inelastic scattering; the meaning of `modest' will be specified in due course. In brief, computation of the spectra involves three steps:
\begin{enumerate}
    \item Calculate, for the given sample's texture, the distribution of its reciprocal lattice vectors in three-dimensional reciprocal space.
    \item Dress every reciprocal lattice vector with two additive kernel functions describing the local distribution (and relative magnitude) of elastic and first-order inelastic scattering intensity.
    \item Sample the resulting three-dimensional dynamic structure factor on those regions of the Ewald sphere accessible in the given experiment.
\end{enumerate}

Rather than formulating the momentum- and energy-resolved scattering cross-section by explicit calculation of the full dynamic structure factor $S(\mathbf{q},\omega)$, we take for our foundation the relatively easily calculable static structure factor $S(\mathbf{q})$, which yields the total (spectrally integrated) inelastic scattering signal, commonly referred to as thermal diffuse scattering (TDS). The present authors have verified in previous studies \cite{Wark2025,Heighway2025} that the model for $S(\mathbf{q})$ described herein successfully predicts the angularly resolved TDS patterns from moderately textured, rolled Cu foils. We then spectrally decompose the total scattering cross-section $[S(\mathbf{q}) \to S(\mathbf{q},\omega)]$ using the principles of energy conservation, detailed balance, and the sum rule. Throughout, we use a highly simplified (linear) phonon dispersion relation, which necessarily compromises the accuracy of the spectra we predict, but keeps the model analytically tractable. We first describe our model for single-crystal scattering, before extending the formalism to arbitrarily textured polycrystals, and finally deriving the meV-IXS spectrum of a perfectly random powder.

\subsection{\label{sec:overview}Static structure factor of a single crystal}

The model of single-crystal thermal diffuse scattering presented in this section is a distillation of work by James\cite{James1948}, Warren\cite{Warren1953,Warren1990}, Paskin\cite{Paskin1958}, and references therein. We provide a more complete exposition in Ref.~\onlinecite{Heighway2025}; here, we focus on the formulae needed to implement the model.

The instantaneous static structure factor of a monatomic group of $N_a$ atoms with time-dependent positions $\{\mathbf{r}_n(t)\}$ at reciprocal-space point $\mathbf{q}$ and time $t$ is
\begin{equation}
    \tilde{s}(\mathbf{q},t)=f^2(\mathbf{q})\left\vert \sum_{n=1}^{N_a} e^{-i\mathbf{q}\cdot\mathbf{r}_n(t)}\right\vert^2\ ,
\end{equation}
where $f(\mathbf{q})$ is the atomic form factor. To calculate the time-averaged structure factor of a single, defect-free, finite-temperature crystallite, $s(\mathbf{q})$, we model each atom's thermal displacement from its equilibrium position $\mathbf{R}_n$ as arising from an equilibrium distribution of noninteracting phonons obeying the same isotropic dispersion relation used in the Debye model of specific heat. The angular frequency of the phonon mode with wavevector $\mathbf{g}$ and polarization state $j$ is assumed to be
\begin{equation}\label{eq:dispersion}
    \omega_{\mathbf{g}j} = \bar{c}|\mathbf{g}|
\end{equation}
regardless of polarization, with $\bar{c}$ being the universal, density-dependent sound speed. The phonons are taken to occupy a spherical Brillouin zone of (reciprocal) radius $g_B$, which, for a cubic crystal, is related to the cubic lattice constant $a$ by
\begin{equation}\label{eq:brillouin-radius}
    g_B = \frac{2\pi}{a}\left(\frac{3N_b}{4\pi}\right)^{\frac{1}{3}}\ ,
\end{equation}
where $N_b$ is the number of atoms in the conventional unit cell. The most energetic phonons therefore have energy
\begin{equation}
    \hbar\bar{c}g_B\equiv k_B\Theta_D\ ,
\end{equation}
where $\Theta_D$ is the density-dependent Debye temperature, as essential parameter governing all phonon structure.

Regardless of the phonon dispersion relation, the time-averaged structure factor may be expressed as a sum of contributions from $\ell$-phonon scattering events (with $\ell=0$ denoting elastic scattering):
\begin{equation}
    s(\mathbf{q}) = s_0(\mathbf{q}) + \sum_{\ell=1}^\infty s_\ell(\mathbf{q})\ .
\end{equation}
The $\ell$-th order scattering cross-section takes the form
\begin{equation}\label{eq:power-series}
    s_\ell(\mathbf{q}) = N_af^2(\mathbf{q})e^{-2M(\mathbf{q})}\frac{(2M)^\ell}{\ell!}C_\ell(\mathbf{q})\ ,
\end{equation}
where $N_a$ is the number of atoms in the crystallite, $M(\mathbf{q})$ is the high-temperature $(T>\Theta_D)$ form of the Debye-Waller factor, which for atoms of mass $m$ reads
\begin{equation}\label{eq:debye-waller}
    M(\mathbf{q}) = \frac{3}{2}\frac{(\hbar q)^2}{mk_B\Theta_D}\frac{T}{\Theta_D}\ ,
\end{equation}
and where the Paskin coefficient $C_\ell(\mathbf{q})$ encodes the degree of structure of $\ell$-th order scattering in reciprocal space, which diminishes at higher orders.

We will restrict our attention to scenarios in which the first-order inelastic scattering dominates that of orders two and above. If we neglect their structure (i.e., assuming $C_\ell\approx1$), the ratio of the second- and first-order inelastic scattering is given directly by the Debye-Waller factor:
\begin{equation}\label{eq:second-to-first}
    \frac{s_2(\mathbf{q})}{s_1({\mathbf{q}})} \approx M(\mathbf{q})\ .
\end{equation}
Hence, there is a joint temperature--scattering-angle domain within which higher-order inelastic scattering may be neglected. To illustrate, for ambient Cu ($m=63.5u$, $\Theta_D=311$~K\cite{Ho1974}, $T=293$~K) probed between its $(200)$ and $(220)$ Bragg peaks ($q\approx2.4g_B$), the second-order inelastic scattering has only about 5\% the strength of the first-order scattering. However, once the 150~GPa Hugoniot state has been reached ($\Theta_D=640$~K, $T\approx3000$~K), that ratio has grown to 20\%\cite{Wark2025}. Probing at larger $q$ rapidly exacerbates this issue, as the Debye-Waller factor grows quadratically with $q$. The transition to the multiphonon scattering regime will be discussed in Sec.~\ref{sec:discussion}.

If we restrict our attention to sufficiently low temperatures and momentum transfers, we can retain only the elastic and first-order inelastic cross-sections:
\begin{equation}
    s(\mathbf{q})\approx s_0(\mathbf{q})+s_1(\mathbf{q})\ .
\end{equation}
To a reasonable approximation, we may conveniently decompose each of these scattering processes into additive contributions from individual reciprocal lattice vectors, such that
\begin{equation}
   s_i(\mathbf{q}) = \sum_j s_i(\mathbf{q}|\mathbf{G}_j)\ .
\end{equation}
The variation of each component of the structure factor in the neighborhood of reciprocal lattice vector $\mathbf{G}$ is
\begin{align}
    \label{eq:sqG_elasic}s_0(\mathbf{q}|\mathbf{G}) &= N_af^2(\mathbf{q})e^{-2M(\mathbf{q})}J(\mathbf{q}-\mathbf{G})\ , \\
    \label{eq:sqG_inelastic}s_1(\mathbf{q}|\mathbf{G}) &= N_af^2(\mathbf{q})e^{-2M(\mathbf{q})}2M(\mathbf{q})W(\mathbf{q}-\mathbf{G})\ .
\end{align}
The functions $J(\mathbf{k})$ and $W(\mathbf{k})$ describe the reciprocal-space distribution of the elastic and first-order inelastic scattering, and are essentially the kernels with which we convolve the reciprocal lattice to construct the full static structure factor. The \emph{shape function} $J(\mathbf{k})$ describes the slight but nonzero delocalization of elastic scattering in reciprocal space owed to the crystal's finite size, and is approximated well by the absolute square of the Fourier transform of the binary function $B(\mathbf{r})$ delineating the volume occupied by the crystal. For example, for a perfect, spherical crystallite of radius $R$, for which
\begin{equation}
    B(\mathbf{r})=\begin{cases}
        1\quad\text{for}\quad0\le r\le R\ ,\\
        0\quad\text{otherwise}\ ,
    \end{cases}
\end{equation}
the shape function obeys, to a reasonable approximation,
\begin{equation}\label{eq:spherical-grain}
    J(\mathbf{k}) \propto \left\vert4\pi R^3\left[\frac{\sin(kR)-kR\cos(kR)}{(kR)^3}\right]\right\vert^2\ .
\end{equation}
To produce the total elastic scattering intensity correctly, the shape function must be normalized such that
\begin{equation}
    \iiint_{\mathbb{R}^3}d^3\mathbf{k}\,J(\mathbf{k})=N_b\left(\frac{2\pi}{a}\right)^3\ .
\end{equation}
Note that, unless it is entirely devoid of defects, the total size of a crystallite may exceed considerably the effective size with which it diffracts. Dislocation structures tend to break grains up into coherently diffracting domains of far smaller size than the grain itself\cite{DeAngelis1995,Dubravina2004}. These coherent domains are the fundamental diffracting bodies in this model, and it is therefore their dimension, rather than that of the grains, that must be encoded into the shape function $J(\mathbf{k})$.

Meanwhile, the isotropic \emph{Warren kernel} $W(\mathbf{k})$ describes the delocalized, slowly varying distribution of first-order inelastic scattering around each reciprocal lattice vector owed to phonons. It takes the form
\begin{equation}\label{eq:warren-kernel}
    W(k) = \begin{cases}
        \frac{1}{3}\frac{g_B^3}{k^2}\quad\text{for}\quad0<k\le g_B\ , \\
        0\quad\quad\ \text{otherwise}\ .
    \end{cases}
\end{equation}
The inverse-square relation reflects the diminishing population of higher-energy phonons towards the edge of each Brillouin zone.

We now go on to describe how to `upgrade' the static structure factor by further discriminating the scattering cross-section by energy transfer.

\begin{figure*}
    \centering
    \includegraphics{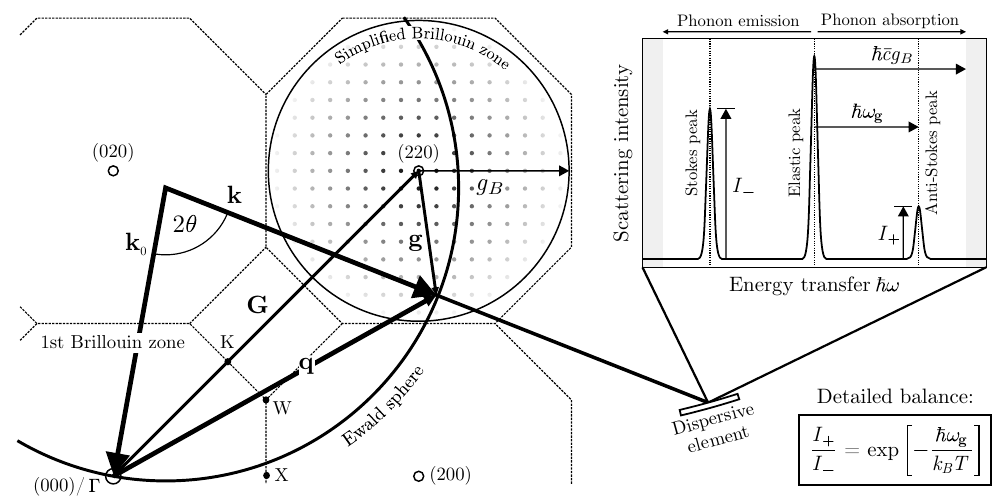}
    \caption{Schematic depiction of first-order (single-phonon) inelastic scattering of a monochromatic, collimated x-ray beam by a single crystal. (Left) Photons of momentum $\hbar\mathbf{k}_0$ are scattered through angle $2\theta$ into momentum state $\hbar\mathbf{k}$ on the Ewald sphere ($|\mathbf{k}|=|\mathbf{k}_0|$) via momentum transfer $\hbar\mathbf{q}$. Scattering occurs with the assistance of phonons within the Brillouin zone centered on the reciprocal lattice vector $\mathbf{G}$ nearest $\mathbf{q}$ with wavevector $\mathbf{g}=\mathbf{q}-\mathbf{G}$. Reciprocal lattice vectors (hollow points) and surrounding Brillouin zones are shown for the reciprocal lattice of a face-centered-cubic (fcc) crystal, with high-symmetry points ($\Gamma$,X,K,W) marked. Solid points mark individual phonon modes within a simplified, spherical Brillouin zone of radius $g_B$. (Right) Photons scattered through wavevector $\mathbf{q}$ may be further discriminated by their energy transfer $\hbar\omega\ll\hbar c|\mathbf{k}|$. Photons either absorb or emit a phonon of energy $\hbar\omega_{\mathbf{g}}=\hbar\bar{c}|\mathbf{g}|$ with relative probability $\exp(-\beta\hbar\omega_{\mathbf{g}})$ set by the principle of detailed balance, leading to the formation of anti-Stokes and Stokes peaks when spectrally dispersed. Also shown is an elastic peak at zero energy transfer arising from the finite crystallite size.}
    \label{fig:monocrystal_scattering}
\end{figure*}

\subsection{Dynamic structure factor of a single crystal}
The static structure factor $s(\mathbf{q})$ is proportional to the total probability of an incident photon of momentum $\hbar\mathbf{k}_0$ being scattered into a new momentum state $\hbar\mathbf{k}$, where $\mathbf{q}=\mathbf{k}-\mathbf{k}_0$, integrated over all possible energy transfers. As described in Sec.~\ref{sec:overview}, the two dominant scattering processes for a single crystallite are generally zero-phonon elastic and single-phonon inelastic scattering, with spectrally unresolved cross-sections $s_0(\mathbf{q})$ and $s_1(\mathbf{q})$, respectively. We seek an expression for the dynamic structure factor $s(\mathbf{q},\omega)$, which expresses the probability of a photon receiving momentum $\hbar\mathbf{q}$ while simultaneously receiving energy $\hbar\omega$. The spectral content of the photons scattered via each process can in fact be derived directly from the static structure factor by simple application of three general constraints on the scattering processes.

Consider the photons scattered through wavevector $\mathbf{q}$ by a single phonon in the vicinity of reciprocal lattice vector $\mathbf{G}$, the geometry for which is shown in Fig.~\ref{fig:monocrystal_scattering}. Scattering of this kind can occur via either the absorption or the emission of a phonon of wavevector $\mathbf{g}=\mathbf{q}-\mathbf{G}$. The energy transferred to or from the photon is necessarily equal to that of the phonon involved, which, in our model, is simply
\begin{equation}
    \hbar\omega_\mathbf{g}=\hbar\bar{c}|\mathbf{q}-\mathbf{G}|\ .
\end{equation}
Moreover, the relative probabilities of phonon absorption $(\omega>0)$ and emission $(\omega<0)$ are fixed by the law of detailed balance, which requires that
\begin{equation}
    s(\mathbf{q},\omega) = e^{-\beta\hbar\omega}s(\mathbf{q},-\omega)\ ,
\end{equation}
where $\beta=(k_BT)^{-1}$. That is, for all finite temperatures, the incident photon is more likely to lose energy than gain it. Together, the principles of energy conservation and detailed balance require that the energy- and momentum-resolved first-order inelastic cross-section assume the form
\begin{equation}
\begin{split}
    \label{eq:deltas}s_1(\mathbf{q},\omega|\mathbf{G}) = A(\mathbf{q}|\mathbf{G})[&e^{-\frac{1}{2}\beta\hbar\omega_{\mathbf{g}}}\delta(\omega-\omega_{\mathbf{g}}) + \\
    &e^{+\frac{1}{2}\beta\hbar\omega_{\mathbf{g}}}\delta(\omega+\omega_{\mathbf{g}})]\ ,
\end{split}
\end{equation}
where $\delta(x)$ is the Dirac delta function. The prefactor $A(\mathbf{q}|\mathbf{G})$ can be deduced immediately from the sum rule, which requires that integrating over all possible energy transfers at a given momentum transfer yield the (known) static structure factor:
\begin{equation}
    \label{eq:sum_rule}\int_{-\infty}^\infty d\omega\,s(\mathbf{q},\omega) = s(\mathbf{q})\ .
\end{equation}
Combining Eqs.~(\ref{eq:deltas}) and (\ref{eq:sum_rule}) with the analytic expression for $s_1(\mathbf{q})$ [Eq.~(\ref{eq:sqG_inelastic})], we find that the probability of momentum transfer $\hbar\mathbf{q}$ and energy transfer $\hbar\omega$ to an incident photon due to single-phonon--assisted scattering from reciprocal lattice vector $\mathbf{G}$ takes the form
\begin{equation}
    s_1(\mathbf{q},\omega|\mathbf{G}) = s_1^+(\mathbf{q},\omega|\mathbf{G}) + s_1^-(\mathbf{q},\omega|\mathbf{G})
\end{equation}
where the functions expressing the up- and down-scattering components of the spectrum are
\begin{widetext}
\begin{equation}\label{eq:dynamic_fs_inelastic}
    s_1^{\pm}(\mathbf{q},\omega|\mathbf{G}) = N_a f^2(\mathbf{q})2M(\mathbf{q})e^{-2M(\mathbf{q})}W\left(\frac{\omega_{\mathbf{g}}}{\bar{c}}\right)f\left(\mp\frac{\hbar\omega_{\mathbf{g}}}{k_BT}\right)\delta(\omega\mp\omega_{\mathbf{g}})\ ,
\end{equation}
\end{widetext}
with
\begin{equation}
    f(x)=\frac{1}{1+e^{-x}}
\end{equation}
being the standard logistic function. The spectrally resolved \emph{elastic} cross-section is trivially evaluated from Eq.~(\ref{eq:sqG_elasic}), since the only possible energy transfer is zero:
\begin{equation}\label{eq:dynamic_sf_elastic}
    s_0(\mathbf{q},\omega|\mathbf{G}) = N_af^2(\mathbf{q})e^{-2M(\mathbf{q})}J(\mathbf{q}-\mathbf{G})\delta(\omega)\ .
\end{equation}

Figure~\ref{fig:monocrystal_scattering} shows an idealised version of the spectrum one might expect to obtain in a typical meV-IXS experiment conducted at an x-ray free-electron laser (XFEL), in which a single crystal is probed over a very narrow $\mathbf{q}$-range, such that essentially only a single phonon mode is accessed. Emission or absorption of a phonon in that mode by the incoming photon leads to the formation of Stokes and anti-Stokes peaks, respectively, whose intensity ratio is set by detailed balance. Coexisting with these inelastic peaks is the elastic peak at zero energy transfer. Note that although the elastic scattering cross-section diminishes far more quickly as one moves away from the reciprocal lattice vectors than does the inelastic, it is also of much greater intrinsic strength; the proportion is, again, given by (twice) the Debye-Waller factor [cf.~Eqs.~(\ref{eq:sqG_elasic},\ref{eq:sqG_inelastic})]. For this reason, the elastic scattering may compete with the inelastic scattering even far from the Bragg condition.

Together, Eqs.~(\ref{eq:dynamic_fs_inelastic}) and (\ref{eq:dynamic_sf_elastic}) yield the additive, leading-order contributions to the dynamic structure factor $s(\mathbf{q},\omega)$ in the vicinity of each reciprocal lattice vector. To calculate the structure factor of a polycrystalline aggregate -- which we will denote by $S(\mathbf{q},\omega)$ to distinguish it from its single-crystal analogue -- one simply needs to know the full distribution of its reciprocal lattice vectors in momentum space.

\subsection{Dynamic structure factor of a textured polycrystal\label{sec:theory-texture}}

\begin{figure*}
    \centering
    \includegraphics{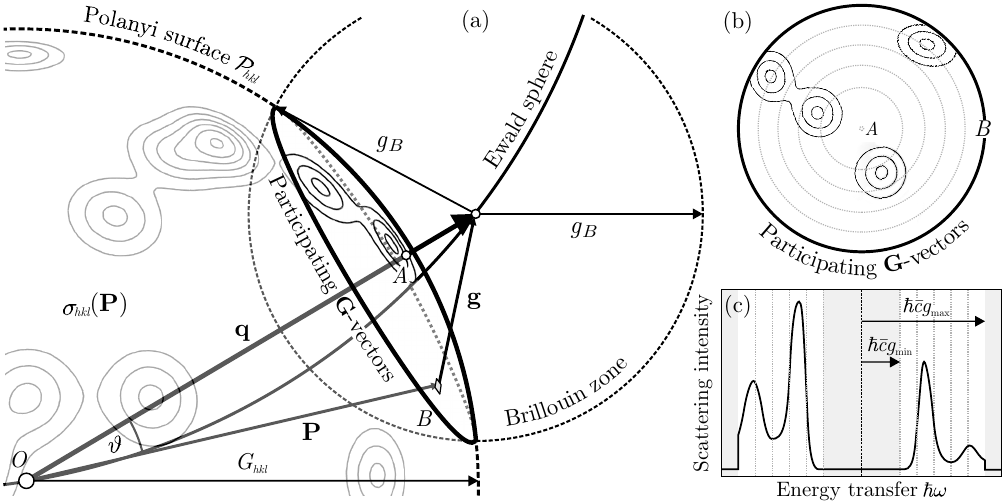}
    \caption{Schematic depiction of first-order (single-phonon) inelastic scattering of a monochromatic, collimated x-ray beam by a polycrystal. (a) An anisotropic distribution of reciprocal lattice vectors in the family $\{hkl\}$ inhabits a spherical Polanyi surface $\mathcal{P}_{hkl}$, with radius $G_{hkl}$ and with $\sigma_{hkl}(\mathbf{P})$ scattering vectors per unit solid angle at reciprocal-space point $\mathbf{P}$. The reciprocal lattice vectors participating in single--phonon-assisted scattering by wavevector $\mathbf{q}$ are those within one Brillouin zone's radius $(g_B)$ of $\mathbf{q}$, and thus form a spherical cap with point of closest approach $A$ and circular boundary $B$. (b) Stereographic view of the cap of participating scattering vectors. Dotted lines are equispaced contours of constant energy transfer. (c) Resulting meV-IXS spectrum with the same energy transfers as in (b) marked by vertical dotted lines. Grey regions mark energy transfers that cannot be mediated by a single phonon for reciprocal lattice vectors on $\mathcal{P}_{hkl}$. Elastic scattering is not pictured.}
    \label{fig:polycrystal_scattering}
\end{figure*}

Under ambient conditions, the reciprocal lattice vectors of a polycrystalline aggregate occupy a set of concentric spheres we term \textit{Polanyi surfaces}. Surface $\mathcal{P}_{hkl}$ comprises all reciprocal lattice vectors from the family $\{hkl\}$, and is characterized by its radius $G_{hkl}=|(hkl)|$ and its multiplicity $j_{hkl}$. As in Ref.~\onlinecite{Heighway2025}, we express the distribution of reciprocal lattice vectors on $\mathcal{P}_{hkl}$ via the number of vectors per unit solid angle, $\sigma_{hkl}$. This angular density necessarily satisfies the normalization condition
\begin{equation}
    \iint_{\mathcal{P}_{hkl}}d\Omega\,\sigma_{hkl} = N_g j_{hkl}\ ,
\end{equation}
where $N_g$ is the total number of grains in the polycrystal, which we will assume are all of identical size and shape.

To calculate the scattering cross-section of the entire polycrystal, we integrate the contributions of every reciprocal lattice vector from every Polanyi surface:
\begin{equation}\label{eq:polanyi_integral}
    S(\mathbf{q},\omega) = \sum_{hkl}\iint_{\mathcal{P}_{hkl}}d\Omega\,\sigma_{hkl}(\mathbf{P})s(\mathbf{q},\omega|\mathbf{P})\ .
\end{equation}
That is, the reciprocal lattice vectors situated at point $\mathbf{P}$ on Polanyi surface $\mathcal{P}_{hkl}$ [of which there are $d\Omega\,\sigma_{hkl}(\mathbf{P})$] each contribute $s(\mathbf{q},\omega|\mathbf{P})$ to the scattering cross-section, where the additive elastic and first-order inelastic components of $s(\mathbf{q},\omega|\mathbf{P})$ are given by Eqs.~(\ref{eq:dynamic_sf_elastic}) and (\ref{eq:dynamic_fs_inelastic}), respectively. The set of Polanyi densities is, in turn, derived from a single orientation distribution function (ODF) that characterizes the polycrystal's texture.

Rather than comprising a pair of symmetrically displaced peaks (as in a single-crystal scattering experiment), the meV-IXS spectrum of a polycrystal will generally comprise a nontrivial distribution of photon energies. The structure of the spectrum may be understood intuitively with reference to Fig.~\ref{fig:polycrystal_scattering}. To calculate the first-order inelastic scattering at reciprocal-space point $\mathbf{q}$, we need only consider those reciprocal lattice vectors from which we can reach $\mathbf{q}$ with the aid of a single phonon wavevector. The locus of points on $\mathcal{P}_{hkl}$ that actively contribute to the first-order inelastic scattering can thus be identified by constructing a Brillouin zone (of radius $g_B$, the greatest allowed phonon wavenumber) centered on $\mathbf{q}$, and finding all points on $\mathcal{P}_{hkl}$ within it. This locus takes the form of a spherical cap. The range of possible energy transfers correlates directly with the range of separations between points on the cap and the point $\mathbf{q}$. The least energetic phonon that can be involved in scattering is that emanating from the reciprocal lattice vector closest to $\mathbf{q}$ [labeled $A$ in Fig.~\ref{fig:polycrystal_scattering}(a)], and has wavenumber $g_{\text{min}}=|q-G_{hkl}|$. The most energetic phonons that can be involved are those emanating from reciprocal lattice vectors on the circumference of the cap [labeled $B$ in Fig.~\ref{fig:polycrystal_scattering}(a)], and have wavenumber $g_{\text{max}}=g_B$. Such extremal reciprocal lattice vectors may or may not actually be present, depending on the texture and on the chosen $\mathbf{q}$-vector. We show in Figs.~\ref{fig:polycrystal_scattering}(b,c) an example of how a simple distribution of participating reciprocal lattice vectors brings about a nontrivial meV-IXS spectrum.

It is crucial to understand, however, that this nontrivial structure exists only at scattering vectors outside the first Brillouin zone $(q>g_B)$, in the so-called \textit{umklapp} scattering regime. Inside the first zone (\textit{normal} scattering), the only reciprocal lattice vector $\mathbf{G}$ from which the scattering vector $\mathbf{q}$ can be reached with the aid of a phonon wavevector $\mathbf{g}$ is $(000)$. That is, only phonon modes for which $|\mathbf{g}|=|\mathbf{q}|$ participate in the scattering process, meaning -- if we invoke the isotropic phonon dispersion relation in Eq.~(\ref{eq:dispersion}) -- that only a single phonon energy is sampled; the meV-IXS spectrum reduces to a pair of Stokes and anti-Stokes peaks once more. Thus, the structure of the spectra is only sensitive to crystallographic texture \emph{outside} the first Brillouin zone, in the umklapp regime.

\begin{figure*}
    \centering
    \includegraphics{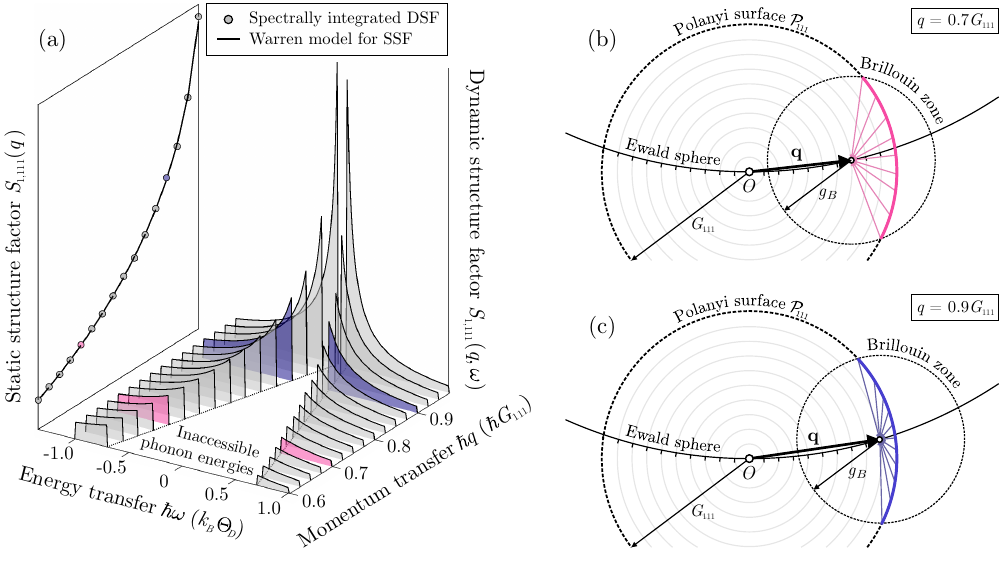}
    \caption{Structure of the first-order (single-phonon) inelastic scattering from a Debye powder, focusing only on contributions from the $\{111\}$ reciprocal lattice vectors of a face-centered-cubic (fcc) crystal. (a) Variation of the dynamic structure factor (DSF) with momentum transfer $\hbar q$, where $q$ is expressed in units of $G_{111}=|(111)|$, as the Bragg condition $(q=G_{111})$ is approached from below. For each spectrum, the energy transfer $\hbar\omega$ is given in units of the maximum phonon energy $k_B\Theta_D$, where $\Theta_D$ is the Debye temperature. Left pane compares the spectrally integrated DSF (gray data points) with the baseline static structure factor (SSF) predicted by the model of Warren (black solid curve). (b,c) Reciprocal-space construction of the scattering condition for two selected spectra at $q = 0.7G_{111}$ and $q=0.9G_{111}$ [colored pink and blue, respectively, in subfigure (a)] illustrating the range of accessible phonons in each case.}
    \label{fig:powder_scattering}
\end{figure*}

\subsection{Dynamic structure factor of an ideal powder}
A powderlike polycrystal is one in which all crystallographic orientations are present and equally probable. For an ideal powder obeying the simple phonon dispersion relation described in Sec.~\ref{sec:overview} (which we will refer to as a \textit{Debye powder}), the single-phonon scattering spectrum can be evaluated analytically. This model spectrum serves as both an excellent illustration of the key physics and as a robust benchmark for any model numerically implementing the texture integral in Eq.~(\ref{eq:polanyi_integral}).

The expression for the up- and down-scattering components of the first-order inelastic scattering is
\begin{equation}
    S_{1}^{\pm}(\mathbf{q},\omega) =\sum_{hkl} \underbrace{\iint_{\mathcal{P}_{hkl}}d\Omega\,\sigma_{hkl}(\mathbf{P})s_1^{\pm}(\mathbf{q},\omega|\mathbf{P})}_{S_{1,hkl}^{\pm}(\mathbf{q},\omega)}\ .
\end{equation}
The natural variables of integration for each Polanyi surface are the polar coordinates $(\vartheta,\varphi)$, defined such that
\begin{equation}
    \mathbf{P}=G_{hkl}(\sin\vartheta\cos\varphi,\sin\vartheta\sin\varphi,\cos\vartheta)\ ,
\end{equation}
where we have defined $\vartheta$ as the angle made by $\mathbf{P}$ with the (fixed) scattering vector $\mathbf{q}$, as in Fig.~\ref{fig:polycrystal_scattering}(a). To exploit the sifting property of the delta functions in the scattering kernels $s_1^{\pm}$, we convert the integral over $\vartheta$ (which expresses the location of a given reciprocal lattice vector) to one over $\omega_{\mathbf{g}}$ (which expresses the energy of the phonon allowing it to contribute to the first-order scattering). Consideration of the geometry in Fig.~\ref{fig:polycrystal_scattering}(a) reveals that
\begin{equation}
    \left\vert\frac{d\vartheta}{d\omega_{\mathbf{g}}}\right\vert\sin\vartheta = \frac{\omega_{\mathbf{g}}}{q\bar{c}^2G_{hkl}}\ .
\end{equation}
Furthermore, for an ideal powder, the angular density of reciprocal lattice vectors on each surface is constant:
\begin{equation}
    \sigma_{hkl}=\frac{N_gj_{hkl}}{4\pi}\ .
\end{equation}
The contribution to the up- and down-scattering from Polanyi surface $\mathbf{P}_{hkl}$ can thus be rendered in the form
\begin{widetext}
\begin{equation}
    S_{1,hkl}^{\pm}(\mathbf{q},\omega) = \frac{1}{6}j_{hkl}N_aN_gf^2(\mathbf{q})2M(\mathbf{q})e^{-2M(\mathbf{q})}\frac{g_B^2}{qG_{hkl}}\int_{\omega_{\mathbf{g}}^{\text{min}}}^{\omega_{\mathbf{g}}^{\text{max}}}d\omega_{\mathbf{g}}\,\frac{1}{\omega_{\mathbf{g}}}f(\mp\beta\hbar\omega_{\mathbf{g}})\delta(\omega_{\mathbf{g}}\mp\omega)\ .
\end{equation}
\end{widetext}
Here, $\omega_{\mathbf{g}}^{\text{min}} = \bar{c}|q-G_{hkl}|$ and $\omega_{\mathbf{g}}^{\text{max}} = \bar{c}g_B$ refer to the lowest and highest energy phonons that contribute to scattering at $\mathbf{q}$, which emerge from points $A$ and $B$ on the spherical cap of participating scattering vectors, respectively. After evaluating both components of the spectrum explicitly, we find that the contribution to the first-order inelastic scattering cross-section from scattering vectors $\{hkl\}$ for a Debye powder is
\begin{widetext}
\begin{equation}\label{eq:S1hkl}
    S_{1,hkl}(\mathbf{q},\omega) = \begin{dcases}
        \frac{1}{6}j_{hkl}N_aN_gf^2(\mathbf{q})2M(\mathbf{q})e^{-2M(\mathbf{q})}\frac{g_B^2}{qG_{hkl}}\frac{1}{|\omega|}f(-\beta\hbar\omega)\quad\text{for}\quad\omega_{\mathbf{g}}^{\text{min}}\le|\omega|\le\omega_{\mathbf{g}}^{\text{max}}\ , \\
        0\quad\text{otherwise}\ .
    \end{dcases}
\end{equation}
\end{widetext}
Note that we may also evaluate analytically these scattering vectors' contribution to the \emph{static} structure factor:
\begin{align}
    S_{1,hkl}(\mathbf{q}) &= \int_{-\infty}^\infty d\omega\,S_{1,hkl}(\mathbf{q},\omega) \\
    &= \frac{1}{6}j_{hkl}N_aN_gf^22Me^{-2M}\frac{g_B^2}{qG_{hkl}}\ln\left(\frac{\omega_{\mathbf{g}}^{\text{max}}}{\omega_{\mathbf{g}}^{\text{min}}}\right) \\
    &= \frac{1}{6}j_{hkl}N_aN_gf^22Me^{-2M}\frac{g_B^2}{qG_{hkl}}\ln\left(\frac{g_B}{|q-G_{hkl}|}\right)\label{eq:warren_SSF}
\end{align}
in exact agreement with Warren's classic expression for the thermal diffuse scattering from a Debye powder\cite{Warren1953}.

In Fig.~\ref{fig:powder_scattering}, we illustrate how the contribution to the meV-IXS spectrum from a single family of reciprocal lattice vectors [Eq.~(\ref{eq:S1hkl})] varies as a function of the momentum transfer $\hbar q$. The partial spectrum generally comprises a pair of asymmetric wings, each of which spans the full range of allowed phonon-mediated energy transfers between $\hbar\omega_{\mathbf{g}}^{\text{min}}$ and $\hbar\omega_{\mathbf{g}}^{\text{max}}$. As the $\mathbf{q}$-vector nears the Polanyi surface (i.e., as the Bragg condition is approached), progressively shorter-wavevector phonons are able to bridge the gap between the cap of participating $\mathbf{G}$-vectors and the $\mathbf{q}$-vector, and thus participate in the scattering. Inclusion of these relatively populous, low-energy phonons causes the total, spectrally integrated cross-section to logarithmically diverge as the Bragg condition is approached, precisely the signature behavior predicted by Warren's model\cite{Warren1953}.

\begin{figure}
    \centering
    \includegraphics{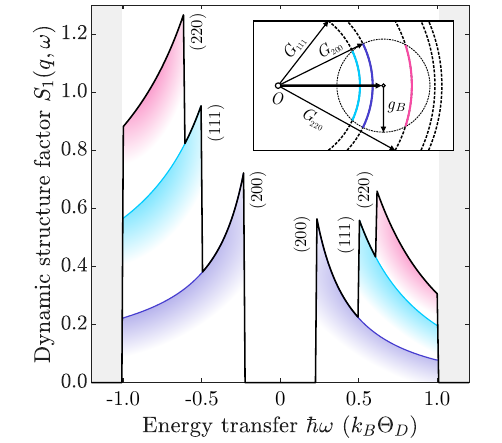}
    \caption{First-order inelastic scattering from a face-centered cubic (fcc) Debye powder evaluated at $q = 1.1G_{200}$, expressed per atom. Contributions from the $\{200\}$, $\{111\}$, and $\{220\}$ scattering vectors are highlighted in blue, cyan, and magenta, respectively. Inset: reciprocal-space construction of the scattering condition. Contributing scattering vectors from each Polanyi surface are those within one Brillouin zone's radius $g_B$ of the scattering vector $\mathbf{q}$.}
    \label{fig:example-spectrum}
\end{figure}

The meV-IXS spectrum from a Debye powder will in general comprise contributions from multiple families of reciprocal lattice vectors simultaneously, each of which is associated with a different range of accessible phonon energies. We show in Fig.~\ref{fig:example-spectrum} a representative spectrum from an fcc Debye powder evaluated at $q=1.11G_{200}$. At this momentum transfer, the $\{111\}$, $\{200\}$ and $\{220\}$ reciprocal lattice vectors are able to contribute to the first-order inelastic scattering; since $\mathbf{q}$ is located nearest $\mathcal{P}_{200}$, the $\{200\}$ vectors contribute over the widest range of energy transfers. For all three families of reciprocal lattice vectors, the greatest energy transfer possible is the Debye energy $k_B\Theta_D=\hbar\bar{c}g_B$.

\section{Simulated spectra from targets for dynamic compression experiments\label{sec:analysis}}

Having built a model for single-phonon inelastic scattering of x-ray photons from linearly dispersed phonons in an arbitrarily textured cubic polycrystal, we will now explore the $q$-dependent structure of the meV-scale scattering spectra one would measure from a typical dynamic-compression target illuminated by an XFEL. Before presenting our calculations, we will comment on the composition of these targets.

\subsection{Target composition\label{sec:targets}}
It is generally accepted that the simplest viable target for a laser-based dynamic-compression experiment comprises (1) a thin ($\sim$10~$\upmu$m) layer of the material of interest and (2) a relatively thick ($\sim50$~$\upmu$m), low-$Z$ (weakly x-ray scattering) ablator situated between the sample layer and the drive laser. The sacrificial ablator conveys multiple benefits. Its main function is to smooth out nonuniformities in the compression wave generated by the laser pulse before they can enter the sample layer\cite{Gorman2022}. A sufficiently thick ablator will also absorb the majority of the soft x-rays radiated by the laser-produced ablation plasma that would otherwise prematurely heat the sample by as much as 10\textsuperscript{3}~K\cite{Preheat}. An optically opaque ablator can further shield the sample from the drive laser's \textit{prepulse}, the relatively low-intensity, stray radiation arriving at the target's surface shortly before the pulse proper that is a common feature of high-power pulsed optical laser systems\cite{Kiriyama2021}. Finally, when the rapidly accelerated ablator impacts the sample, the resulting transfer of mechanical energy often brings about a higher pressure in the latter than if it were irradiated directly by the drive laser\cite{Forbes}, widening considerably the range of thermodynamic states accessible using a drive laser of a given power. For these reasons, we consider the ablator to be indispensable. Its presence in any viable target will therefore be taken as read. We will not consider additional components commonly found in dynamic-compression targets, such as flash coatings, heat shields, or velocimetry windows.

\begin{figure*}
    \centering
    \includegraphics{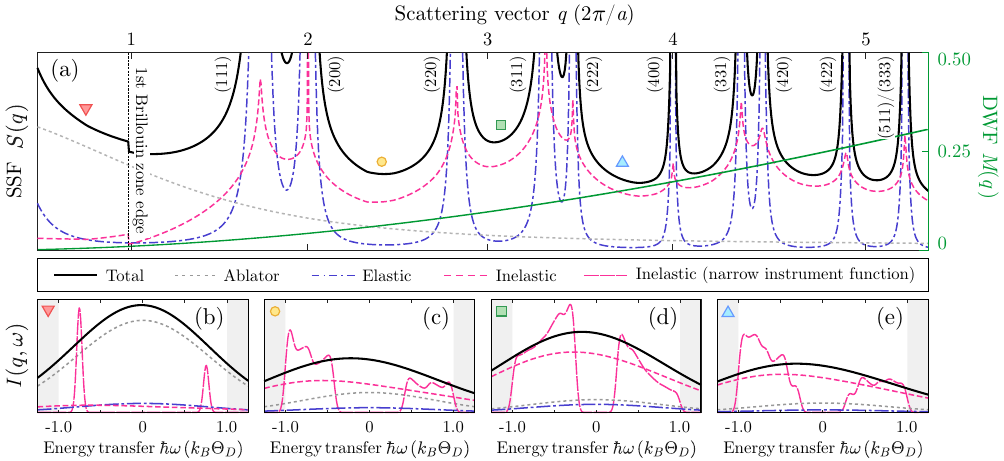}
    \caption{Overview of the structure factor of a representative dynamic-compression target comprising a 50~$\upmu$m structureless plastic ablator and 25~$\upmu$m of powderlike copper (Cu) at 293~K. (a): dependence of the static structure factor (SSF) $S(q)$ on scattering vector $q$, decomposed into contributions from the (coherent) ablator scattering, the elastic scattering from the Cu, and the first-order inelastic scattering from the Cu. Overlaid is the variation of the Debye-Waller factor (DWF) $M(q)$ [Eq.~(\ref{eq:debye-waller})]. Bragg peaks are labeled by their Miller indices. The edge of the first (spherical) Brillouin zone is marked with a dashed vertical line at $g_B\approx0.98(2\pi/a)$ [Eq.~(\ref{eq:brillouin-radius})]. (b-e): dynamic structure factor (DSF) $S(q,\omega)$ at four selected $q$-values (one inside the first Brillouin zone, three outside) marked by colored symbols in (a). The total and individual DFSs have been convolved with a Gaussian instrument function (IF) with a full width at half maximum (FWHM) of 50~meV to yield a simulated intensity $I(q,\omega)$. Also shown are the inelastic scattering spectra from the Cu calculated with a 2~meV FWHM IF to better expose the underlying structure.}
    \label{fig:q-overview}
\end{figure*}

For targets that must be fabricated in large numbers (hundreds or perhaps even thousands for an meV-IXS experiment\cite{Descamps2020a}), the most practical choice of ablator material is usually a polymer. Such ablators (common examples of which include Kapton-B and Parylene-N) are inexpensive, easy to handle, and suitably low in atomic number. However, their lack of long-range atomic order means that, when they are illuminated by an XFEL, polymer ablators will produce largely structureless scattering that coexists with scattering from the sample layer. Though extremely weak compared with a crystalline sample's Bragg peaks, diffraction from the ablator is easily of sufficient strength to compete with the sample's inelastic scattering, as we will show. Our first objective will be to quantify this competition.

Throughout, our prototypical target will be a 25-$\upmu$m-thick copper (Cu) foil bonded to a 50-$\upmu$m-thick Kapton-B ablator, the same targets used in previous experimental studies of TDS detailed in Refs.~\onlinecite{Wark2025,Heighway2025}. We model the Cu ($\rho_{\text{Cu}} = 8.96$~g\,cm\textsuperscript{$-3$})\cite{CRC} as an fcc powder with lattice constant $a$ = 3.615~\AA\cite{Arblaster}, Debye temperature $\Theta_D$ = 311~K\cite{Ho1974} and Debye energy $k_B\Theta_D$ = 26.8~meV at ambient conditions. For the elastic component of the Cu scattering, $S_0(\mathbf{q})$, we use a monotonically decreasing, computationally efficient approximation of the shape function for a spherical grain described by Eq.~(\ref{eq:spherical-grain}) (see Supplementary Material). We use a coherently scattering domain size of $2R$ = 30~nm, a figure appropriate to heavily cold-worked metallic polycrystals\cite{Ungar2001,Zhu2003,Ungar2005} that we have found reproduces the diffraction-peak linewidth of machine-rolled Cu foils well in previous work\cite{Heighway2025}. We model the polyimide ablator ($\rho_{\text{abl}} = 1.42$~g\,cm\textsuperscript{$-3$})\cite{Kapton2013} as a completely amorphous form of H\textsubscript{10}C\textsubscript{22}N\textsubscript{2}O\textsubscript{5}\cite{Kapton2013} (that is, with zero spatial correlation between similar or dissimilar atoms) such that its dynamic structure factor reduces to
\begin{equation}
    S_{\text{abl}}(\mathbf{q},\omega) = \sum_i N_i f_i^2(\mathbf{q})\delta(\omega)\ ,
\end{equation}
where $N_i$ is the number of atoms of element $i$ in the ablator and $f_i(\mathbf{q})$ is their atomic form factor. Treating the ablator as an ideal gas is reasonable here: what little structure exists in the static structure factor of Kapton-B is restricted to $q \lesssim 1.4(2\pi/a)$\cite{Wark2025}; any structure above this scattering vector is weak and largely insensitive to shock pressure. We further assume that, as the polymer ablator is amorphous, its elastic scattering component dominates its inelastic\cite{Descamps2020a}. All atomic form factors are taken from calculations by Cromer and Mann\cite{Cromer1968} fitted to the Mott-Bethe form by Thorkildsen\cite{Thorkildsen2023}.

For both the Cu sample and the polyimide ablator, we neglect the Compton (incoherent) contribution to the scattering, on the basis that the Compton shift suffered by hard x-rays will typically be of order 100~eV, many orders of magnitude greater than the spectral range over which the meV-IXS spectra would be recorded. We do not attempt to incorporate extrinsic experimental effects into our calculations (such as beam polarization, self-absorption of x-rays by the target assembly, or attenuation by detector filters) so as to `decouple' our predictions from any one particular experimental configuration and retain transferability.

\subsection{Scattering from a powder target}
Our exploration of simulated meV-IXS spectra from the prototypical dynamic-compression targets described in Sec.~\ref{sec:targets} centers around Fig.~\ref{fig:q-overview}. In this first instance, we treat the Cu sample as textureless.

We plot in Fig.~\ref{fig:q-overview}(a) the coherent scattering from both the 50-$\upmu$m-thick polyimide ablator (gray) and from the 25-$\upmu$m-thick Cu sample, with the latter decomposed into its elastic (blue) and first-order inelastic (pink) components. The cross-sections are shown at the scale of the inelastic scattering, whose typical strength is at least two orders of magnitude less than that of the elastic (Bragg) peaks. We also show for reference the Debye-Waller factor $M(q)$ for the Cu sample under these ambient conditions (green), whose value expresses the intensity of the second-order inelastic scattering [relative to the first order, per Eq.~(\ref{eq:second-to-first})] which our model is omitting.

It is immediately apparent that at scattering vectors inside the first Brillouin zone, single-phonon inelastic scattering from the mid-$Z$ Cu layer is many times weaker than elastic scattering from the low-$Z$, amorphous ablator. Outside the first zone, however, the situation is reversed; as one moves to higher $q$-values, the ablator scattering steadily decreases, while the inelastic scattering rapidly increases (initially, at least) owing to the prevailing factor of $2M(\mathbf{q})$ to which it is directly proportional [see Eq.~(\ref{eq:S1hkl})]. We see that at $q$-values midway between the first pair of widely separated Bragg peaks [(200) and (220)], the first-order inelastic scattering is already dominant, while still representing 95\% of the total inelastic scattering [per Eq.~(\ref{eq:power-series})]. On this basis alone, there is a compelling case for moving single-phonon meV-IXS measurements outside the first Brillouin zone.

To underscore this point, we show in Figs.~\ref{fig:q-overview}(b-e) synthetic spectra at four different momentum transfers, one within the first zone, three without. The spectra are calculated by convolving the dynamic structure factor with a Gaussian instrument function (IF), $\Lambda(\omega)$:
\begin{equation}\label{eq:convolution-abstract}
    I(\mathbf{q},\omega) = S(\mathbf{q},\omega)*\Lambda(\omega)
\end{equation}
We use an IF of 50 meV width, which is representative of the resolution attainable using the state-of-the-art meV-IXS platform at EuXFEL\cite{McBride2018,Descamps2020a,Wollenweber2021} (cf.\ $k_B\Theta_D$ = 26.8~meV). Within the first zone, the inelastic signal comprises a pair of (overlapping) Stokes and anti-Stokes peaks, as expected in the normal scattering regime. However, their contribution to the total signal is overwhelmed by the broad elastic peak from the ablator centered at zero energy transfer; the resulting signal thus takes a unimodal and almost perfectly symmetric form that is largely insensitive to the Cu sample's temperature. By contrast, the total dynamic structure factor outside the first zone is increasingly dominated by the sample's inelastic scattering, allowing the signature negative skew of the meV-IXS spectrum to re-emerge.

Evidently, stepping outside the first zone provides a means of escaping the pervasive ablator scattering that threatens to swamp the inelastic spectrum of interest. The price one pays is that the spectrum becomes far more complex. An attractive feature of meV-IXS performed in the normal scattering regime is that the inelastic signal comprises only two peaks, which are relatively easily fitted to a pair of semi-empirical Lorentzian functions whose intensity ratio is set by detailed balance\cite{Descamps2020a}. As can be seen from Figs.~\ref{fig:q-overview}(b-e), in which we also plot the Cu inelastic scattering with a very narrow (2~meV) Gaussian IF, the intermediate-$q$ spectra take on a nontrivial, multimodal structure, for the reasons given in Sec.~\ref{sec:theory-texture}.

\begin{figure*}
    \centering
    \includegraphics{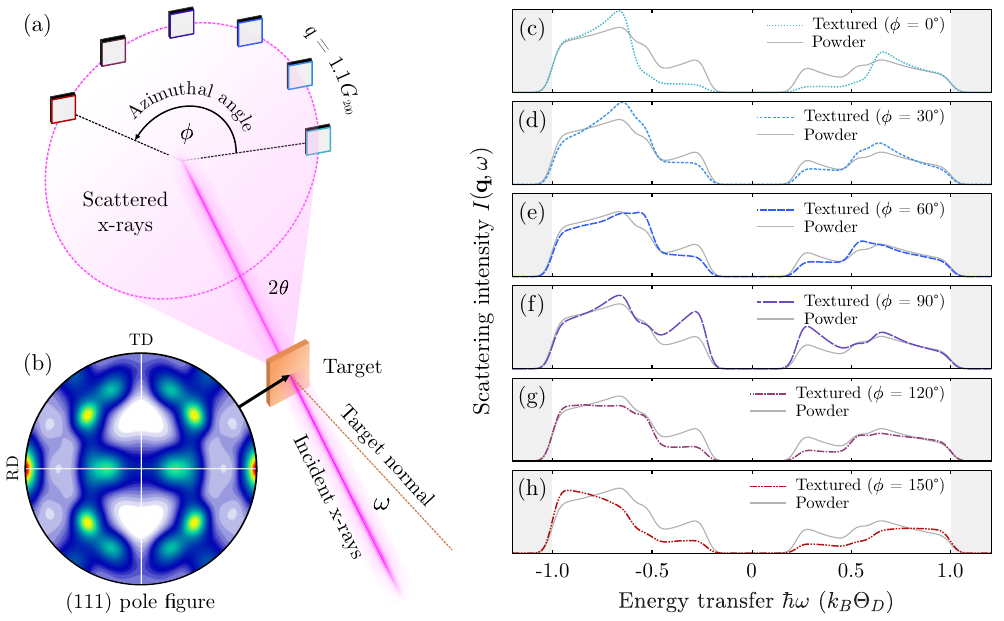}
    \caption{Sensitivity of meV-IXS spectra in the umklapp scattering regime to the crystallographic texture of a machine-rolled copper target. (a) Simulation geometry. The incident x-ray beam travels in the plane defined by the target normal direction (ND) and its rolling direction (RD), at angle $\omega = 22.5^\circ$ to the ND. Pointlike detectors are placed at a fixed scattering vector of $q = 1.11G_{200}$ and at azimuthal angles $\phi$ separated by $30^\circ$ around the incident beam direction, where $\phi=0^\circ$ corresponds to scattering into the ND/RD plane away from the ND. (b) Equal-area pole figure showing the distribution of (111) plane normals for the simulated copper target. (c-h) Variation of first-order meV-IXS spectra with azimuthal angle $\phi$, calculated using a narrow (2~meV width) instrument function. Spectra obtained from a powder (which, by symmetry, are insensitive to $\phi$) are superimposed for comparison.}
    \label{fig:q-texture}
\end{figure*}

At first sight, it might appear that the umklapp spectra could be fitted to a model in a manner similar to the Stokes and anti-Stokes peaks of the normal spectra. In effect, only three variables are required to calculate the meV-IXS spectrum from a Debye powder, according to Eq.~(\ref{eq:S1hkl}): the lattice constant $a$, which is easily measured independently using \textit{in situ} x-ray diffraction; the Debye temperature $\Theta_D$, whose value constrains the total extent of the inelastic spectrum; and the temperature $T$, which controls its skewness. Thus, it would be possible in principle to fit the umklapp spectra directly and thus obtain a joint $(\Theta_D,T)$ distribution. If one could further claim to know the density-dependent Debye temperature (from first-principles modeling, for example), the temperature would be the sole remaining fitting parameter.

However, such a simplistic fit is predicated on the sample being a perfect powder. Very rarely do specimens prepared by machine rolling or physical vapor deposition form without some degree of preferential crystallographic orientation. As noted in Sec.~\ref{sec:theory-texture}, such texture will radically alter the spectral structure. We now go on to quantify the degree of structure for a representative dynamic-compression target.

\subsection{Coupling of spectra to texture}

The orientation distribution function (ODF) that we will use to illustrate the dependence of meV-IXS spectral structure on crystallographic texture is that of machine-rolled, 25-$\upmu$m-thick Cu foils whose thermal diffuse scattering patterns we have previously analyzed\cite{Wark2025,Heighway2025}. These foils exhibit a characteristic fcc rolling texture comprising particularly strong copper, Goss, and cube components [with Euler angles $(\varphi_1,\Phi,\varphi_2)$ of $(90,35,45)^\circ$, $(0,45,90)^\circ$, and $(0,0,0)^\circ$, respectively]. For simplicity, we assume that the sample has perfect mirror symmetry in the rolling, transverse, and normal directions.

Our experimental configuration is shown in Fig.~\ref{fig:q-texture}(a). The x-ray beam meets the Cu sample normal at a non-normal incidence angle of $\omega=22.5^\circ$ (the better to demonstrate the anisotropy of the sample grains' orientation distribution) and travels in the plane defined by the normal and rolling directions. We calculate meV-IXS spectra at a set of momentum transfers of common magnitude $q=1.11G_{200}$, but with equispaced scattering directions around the incident beam path, denoted by their azimuthal angle $\phi$. At this momentum transfer, the $\{111\}$, $\{200\}$, and $\{220\}$ reciprocal lattice vectors participate in the phonon-assisted scattering process (as for the powder spectrum shown in Fig.~\ref{fig:powder_scattering}).

In Figs.~\ref{fig:q-texture}(c-h), we plot an aggregate of first-order meV-IXS spectra from the textured Cu sample calculated at several azimuthal angles between $0^\circ$ and $150^\circ$ inclusive. At each $\mathbf{q}$-value, a different subset of the anisotropically distributed reciprocal lattice vectors participates in the phonon-assisted scattering process, per the geometric construction in Fig.~\ref{fig:polycrystal_scattering}. While the total range of allowed energy transfers is invariant (the limits taking values $\hbar\bar{c}g_{\text{min}} = \hbar\bar{c}|q - G_{200}|\approx0.23k_B\Theta_D$ and $\hbar\bar{c}g_{\text{max}} = k_B\Theta_D$), the relative probability of any given phonon energy being sampled changes dramatically as the caps of participating scattering vectors sweep around the Polanyi surfaces. The sample's crystallographic texture is thus baked directly into its meV-IXS spectra.

The sensitivity of the dynamic structure factor to texture would appear to put working in the umklapp scattering regime at a huge disadvantage. Within the model used here, at least, this coupling does \emph{not} manifest in the normal scattering regime. For scattering vectors inside the first Brillouin zone $(q<g_B)$, the only reciprocal lattice vector $\mathbf{G}$ whence the scattering vector $\mathbf{q}$ can be reached with the aid of a single phonon wavevector $\mathbf{g}$ is $(000)$; since the $(000)$ scattering vectors are concentrated at a single, static point whose density depends only on the total number of grains in the polycrystal, their collective contribution to the inelastic scattering has no dependence on texture. To fit the isolated pair of Stokes and anti-Stokes peaks measured in the first Brillouin zone thus requires no knowledge of the sample's ODF, the unambiguous determination of which usually requires an extensive, \textit{ex situ} mapping using electron backscatter diffraction (EBSD). To characterize the texture of the large volume of material required to realize an meV-IXS--based temperature measurement with sufficient confidence to forward-model their umklapp spectra is no small task.

Fortunately, there exists a methodology owed to Dornheim \textit{et al.}\ whereby the temperature of an arbitrarily complex system may be extracted from its dynamic structure factor in a manner that is completely agnostic to its origin or functional form\cite{Dornheim2022,Dornheim2023}, allowing us to circumvent the effects of texture entirely. We will briefly recap the structure of Dornheim's formalism before demonstrating its application to the umklapp meV-IXS spectra of arbitrarily textured polycrystals.

\subsection{\label{sec:laplace}Temperature extraction in the Laplace domain}
Consider the scattering spectrum of a dynamically compressed polycrystal measured at a particular momentum transfer $\hbar\mathbf{q}$ as a function of energy transfer $\hbar\omega$, resulting from the convolution of its dynamic structure factor (DSF) $S(\mathbf{q},\omega)$ and a normalized instrument function (IF) $\Lambda(\omega)$, as in Eq.~(\ref{eq:convolution-abstract}):
\begin{equation}\label{eq:convolution-specific}
    I(\mathbf{q},\omega) = \int_{-\infty}^\infty d\omega'\,S(\mathbf{q},\omega')\Lambda(\omega-\omega')\ .
\end{equation}
We introduce the two-sided Laplace transform of the scattering intensity, which we define here as
\begin{equation}
    \mathcal{L}[I(\mathbf{q},\omega)] = \int_{-\infty}^\infty d\omega\,I(\mathbf{q},\omega) e^{\omega\tau}\ ,
\end{equation}
and is a function of the scattering vector $\mathbf{q}$ and the ancillary timelike variable $\tau$. Since the Laplace transform obeys the convolution theorem, the transforms of the measured spectral intensity and of the underlying DSF are simply related by
\begin{align}
    \frac{\mathcal{L}[I(\mathbf{q},\omega)]}{\mathcal{L}[\Lambda(\omega)]} &= \mathcal{L}[S(\mathbf{q},\omega)]\ \\\label{eq:L-ratio}
    &= F(\mathbf{q},\tau)\ .
\end{align}
We now examine further the  transform of the DSF, $F(\mathbf{q},\tau)$. The law of detailed balance requires that, for a system in thermodynamic equilibrium,
\begin{equation}\label{eq:detailed-balance}
    S(\mathbf{q},\omega) = e^{-\beta\hbar\omega}S(\mathbf{q},-\omega)\ .
\end{equation}
Separating the Laplace transform of $S(\mathbf{q},\omega)$ into two integrals over positive and negative domains and inserting Eq.~(\ref{eq:detailed-balance}) into the latter's integrand allows us to recast it as
\begin{align}
    F(\mathbf{q},\tau) &= \int_0^\infty d\omega\,S(\mathbf{q},\omega)e^{\omega(\beta\hbar-\tau)} + \int_0^\infty d\omega\,S(\mathbf{q},\omega)e^{\omega\tau} \\
    &= F(\mathbf{q},\beta\hbar-\tau)\ .\label{eq:F-symmetry}
\end{align}
Taking the derivative of Eq.~(\ref{eq:F-symmetry}) with respect to its second argument immediately yields
\begin{equation}
    F'(\mathbf{q},\tau)=-F'(\mathbf{q},\beta\hbar-\tau)\ ,
\end{equation}
whence it follows that $F'(\mathbf{q},\beta\hbar/2)=0$. This is to say that the Laplace transform of the dynamic structure factor necessarily possesses an extremum whose location
\begin{equation}\label{eq:laplace-minimum}
    \tau_D=\frac{\hbar}{2k_BT}
\end{equation}
uniquely identifies the thermodynamic temperature $T$; the physical origin of the DSF is immaterial. Hence, according to Eq.~(\ref{eq:L-ratio}), if we Laplace transform the meV-IXS spectrum of a dynamically compressed polycrystal, and divide the result by the transform of the (presumed known) IF, the extremum of the resulting function allows us to determine its temperature without needing to know anything about its crystallographic texture (or, indeed, its Debye temperature).

\begin{figure*}
    \centering
    \includegraphics{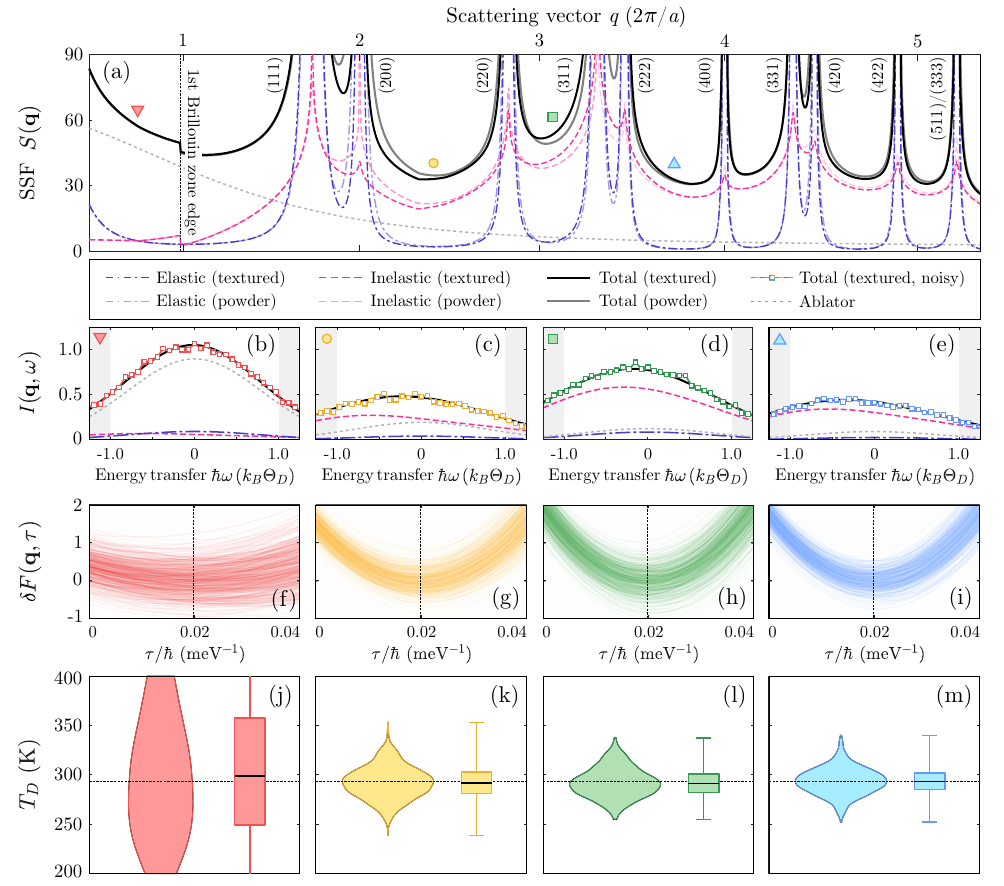}
    \caption{Temperature extraction from a 25-$\upmu$m-thick, textured, ambient (293~K) copper (Cu) foil adhered to 50~$\upmu$m of structureless plastic ablator material, using Dornheim \textit{et al}.'s Laplace-transform--based method\cite{Dornheim2022}. (a) dependence of the static structure factor (SSF) $S(\mathbf{q})$ on the magnitude of the scattering vector $q$ at a fixed azimuthal angle $\phi$, decomposed into contributions from the (coherent) ablator scattering, the elastic scattering from the Cu, and the first-order inelastic scattering from the Cu, all expressed per atom. Components of the SSF for an ideal powder are shown by lighter curves. Bragg peaks are labeled by their Miller indices. The edge of the first (spherical) Brillouin zone is marked with a dashed vertical line at $g_B = 0.98(2\pi/a)$ [Eq.~(\ref{eq:brillouin-radius})]. (b-e) dynamic structure factor (DSF) $S(\mathbf{q},\omega)$ at four selected $\mathbf{q}$-values (one inside the first Brillouin zone, three outside) marked by coloured symbols in (a). The total and individual DSFs have been convolved with a Gaussian instrument function (IF) with a full width at half maximum (FWHM) of 50~meV to yield a simulated intensity $I(\mathbf{q},\omega)$. Also shown are examples of `artificial' spectra, sampled in bins of 2~meV width and with Poissonian noise of a consistent level added. (f-i) Aggregated IF-corrected Laplace transforms of 500 randomly generated artificial spectra, offset by their average minimal value [$\delta F(\mathbf{q},\tau) = F(\mathbf{q},\tau) - \langle F(\mathbf{q},\tau_D)\rangle$ where $\tau_D = \arg \min_{\tau \in \mathbb{R}} F(\mathbf{q},\tau)$]. Dashed vertical lines mark the expected minimum location at $\tau = \hbar/(2k_BT)$, where $T = 293$~K. (j-m) Violin plots and box-and-whisker plots showing the distribution of temperatures derived from the minima of the Laplace transforms of $I(\mathbf{q},\omega)$. Dashed horizontal lines indicate the true temperature of 293~K.}
    \label{fig:temperature-fitting}
\end{figure*}

Note that the measured meV-IXS spectrum of our modelled multilayered targets will in reality comprise both elastic and inelastic scattering from the crystalline sample, as well as elastic scattering from the ablator:
\begin{equation}
    S(\mathbf{q},\omega) = S_{\text{abl}}(\mathbf{q})\delta(\omega) + S_0(\mathbf{q})\delta(\omega) + S_1(\mathbf{q},\omega)\ .
\end{equation}
Since the Laplace transform of the Dirac delta function is unity, the decomposition of the transform of the full target's DSF is
\begin{equation}
    \frac{\mathcal{L}[I(\mathbf{q},\omega)]}{\mathcal{L}[\Lambda(\omega)]} = S_{\text{abl}}(\mathbf{q}) + S_0(\mathbf{q}) + F_1(\mathbf{q},\tau)\ ,
\end{equation}
where $F_1 = \mathcal{L}[S_1]$. Therefore, at a given momentum transfer $\hbar\mathbf{q}$, the effect of elastic scattering is simply to add a constant offset to $\mathcal{L}[I]/\mathcal{L}[\Lambda]$; its minimum is still that of $F_1$, whose abscissa $\tau_D$ yields the temperature of the crystalline sample layer only via Eq.~(\ref{eq:laplace-minimum}). This is to say that the presence of stray elastic scattering does not impair of the accuracy of the temperature measurement. It will, however, degrade its \emph{precision}. Assuming the measured meV-IXS spectra are governed by Poissonian statistics, a stronger elastic scattering component brings about greater noise, which will in real terms diminish the signal-to-noise ratio of the inelastic signal of interest. Here, we see the merit of probing at intermediate $q$-values where the ablator scattering is weaker.

To test the efficacy of the Laplace-transform method, we calculate at four different $\mathbf{q}$-values the distribution of temperatures extracted from ensembles of artificial meV-IXS spectra generated from a textured dynamic-compression target. The four momentum transfers considered have the same $q$ ($2\theta$) values as those whose spectra are pictured in Fig.~\ref{fig:q-overview}, and have an (arbitrarily chosen) azimuthal angle of $\phi=150^\circ$. We generate the meV-IXS spectra by taking the dynamic structure factor $S(\mathbf{q},\omega)$ convolved with a 50-meV instrument function $\Lambda(\omega)$ as usual, discretizing it into bins of 2~meV width, and adding uncorrelated, purely Poissonian noise to the intensity of each bin. The absolute level of the noise (whose value is in practice set by the number of spectra one is willing or able to gather and average over) is not essential for our purposes; it is only important that its value is calculated consistently across all spectra. Here, we choose a noise level such that $I(\mathbf{q},\omega)=1$ corresponds to $10^3$ `counts', and therefore is expected to experience shot-to-shot variation of order 3\%.

The results of the temperature extraction are shown in Fig.~\ref{fig:temperature-fitting}. The textured target's total static structure factor and its various components are plotted in Fig.~\ref{fig:temperature-fitting}(a). We observe that the structure of the scattering pattern between the Bragg peaks is remarkably similar to that of an ideal powder, despite the sample's moderate texture. This, as explored at length in Ref.~\onlinecite{Heighway2025}, is owed to the characteristically delocalized and slowly varying form of the Warren kernel $W(k)$ [Eq.~(\ref{eq:warren-kernel})] dressing each reciprocal lattice vector. We show examples of noisy meV-IXS spectra evaluated at each of the four chosen $\mathbf{q}$-values in Figs.~\ref{fig:temperature-fitting}(b-e), and in Figs.~\ref{fig:temperature-fitting}(f-i), we plot the IF-corrected Laplace transforms of 500 such randomly generated spectra. We see at once the contrast in behavior at low and high $q$: while the distribution of $F(\mathbf{q},\tau)$ curves for the three scattering vectors outside the first Brillouin zone is tightly clustered, that of the low-$q$ spectra distribution is comparatively diffuse. When we convert the minima of these distributions into temperatures [using Eq.~(\ref{eq:laplace-minimum})], the difference is striking. At the $\mathbf{q}$-point just inside the first Brillouin zone, the distribution of fitted temperatures is extremely broad, having an interquartile range of about 100~K and a full range spanning from 150 to 600~K. By contrast, the distributions for the three $\mathbf{q}$-points outside the first zone have interquartile ranges of just 20~K. Thus, by probing at scattering vectors beyond the first zone, we suppress the additional noise caused by ablator scattering, and vastly increase the precision of the temperature deduced from the meV-IXS spectra.

Crucially, we see that, at every $\mathbf{q}$-value, the median inferred temperature does indeed coincide with the true temperature of 293~K with which the artificial spectra were generated. We reiterate that no knowledge whatsoever of the sample layer's underlying crystallographic texture was required. The apparent barrier represented by the texture sensitivity and relatively complex structure of meV-IXS spectra measured in the umklapp regime is therefore easily surmounted by Dornheim \textit{et al.}'s Laplace-transform approach\cite{Dornheim2022,Dornheim2023}.

\section{\label{sec:discussion}Discussion}
Our main argument may be summarized as follows. The nature of dynamic-compression experiments necessitates the use of an ablator. If said ablator is made of a largely structureless polymer (as is often the case), its scattering may be strong enough at low $q$-values to swamp meV-IXS from the polycrystalline sample whose temperature one means to measure, rendering the first Brillouin zone `uninhabitable'. By moving to higher momentum transfers in order to evade the ablator scattering (thus moving from the normal to the umklapp scattering regime), we instead put ourselves at the mercy of crystallographic texture, the details of which now register directly in the structure of the meV-IXS spectrum and are nontrivial to forward-model. However, this additional complexity can be circumvented by appealing to the law of detailed balance, which allows us to extract the sample temperature from its spectrum (assuming the instrument function is sufficiently well constrained) without needing any knowledge of its orientation distribution function. For these reasons, we believe that escaping the bounds of the first Brillouin zone is a viable (and perhaps even necessary) step in the progression of meV-IXS--based thermometry for dynamically compressed matter. Here, we discuss our model and proposed strategy more generally and highlight profitable routes for further research.

\subsection{Multiphonon scattering}
Throughout our study, we have neglected higher-order (multiphonon) inelastic scattering. For the ambient Cu modeled here, this omission is justified: according to Eq.~(\ref{eq:power-series}), the total fraction of the inelastic scattering that single-phonon scattering events constitute is at least 90\% even at scattering vectors three Brillouin zones' radii ($3g_B=5.1$~\AA\textsuperscript{$-1$}) from the origin of reciprocal space [ignoring the structure encoded by the Paskin coefficients $\{C_\ell(\mathbf{q})\}$]. Provided one probes at momentum transfers below this threshold, the relatively simple first-order scattering model outlined in Sec.~\ref{sec:theory} in applicable. Indeed, in this regard, there exists a `Goldilocks zone' in $\mathbf{q}$-space, comprising scattering vectors large enough that ablator scattering is not overwhelming, and small enough that multiphonon scattering is negligible. For ambient Cu, the basin between the $(200)$ and $(220)$ Bragg peaks represents such a zone.

\begin{figure}
    \centering
    \includegraphics{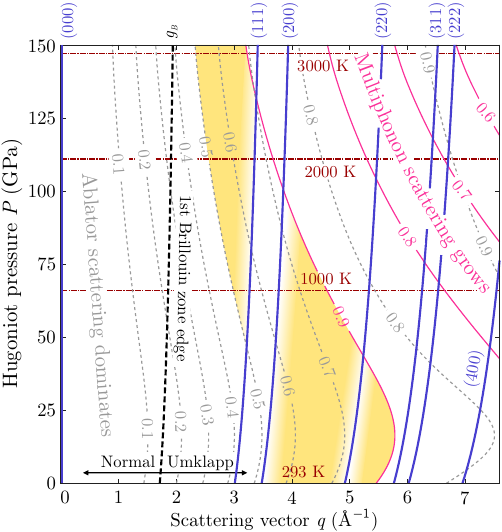}
    \caption{Map of the strength of single-phonon inelastic x-ray scattering relative to other diffuse scattering sources for a target comprising a 50 $\upmu$m polyimide ablator and 25~$\upmu$m Cu powder measured on the Hugoniot. Solid pink contours mark the fraction of the total inelastic scattering that the single-phonon scattering constitutes. Dashed gray contours mark the fraction of summed inelastic and ablator scattering that the inelastic scattering constitutes. Solid blue lines mark Bragg peaks. Dashed black line marks the first Brillouin zone edge. Dot-dashed red lines mark Hugoniot temperatures. The gold region marks a potentially desirable $q$-range over which the single-phonon inelastic scattering is at least half as strong as the ablator scattering, constitutes at least 90\% of the total inelastic scattering, and is unobscured by Bragg scattering. Inelastic scattering cross-sections are calculated assuming no structure beyond that imparted by the atomic form and Debye-Waller factors [i.e., $C_\ell(\mathbf{q})=1$, see Eq.~(\ref{eq:power-series})]. Hugoniot states are calculated using SESAME EOS 3336\cite{SESAME}.}
    \label{fig:goldilocks}
\end{figure}

However, during dynamic compression, the bounds of this habitable zone will shift considerably. We show schematically how the single-phonon--dominated region evolves as a function of shock pressure for our polyimide-copper targets in Fig.~\ref{fig:goldilocks}, using Hugoniot states calculated from SESAME equation of state 3336\cite{SESAME}. For weak shocks ($P<20$~GPa), the habitable zone shifts to slightly higher momentum transfers. This is because the shock-compression-induced increase in the steepness of the potential well confining each atom (expressed by the Debye temperature $\Theta_D$) initially outpaces the shock-heating-induced increase in its thermal energy (expressed by the temperature $T$), such that the Debye-Waller factor $M(\mathbf{q})$ [Eq.~\ref{eq:debye-waller}] decreases, and the inelastic scattering \emph{weakens}. However, beyond $25$~GPa where shock heating rapidly comes to dominate and the inelastic scattering cross-section grows, the habitable zone migrates to lower $q$-values. We see that at a Hugoniot pressure of 150~GPa, it may in fact be prudent to abandon the basin between the $(200)$ and $(220)$ Bragg peaks (which starts to admit appreciable multiphonon scattering) in favor of the region between the $(111)$ peak and the first Brillouin zone edge (where the ablator scattering is now much weaker relative to the increased inelastic scattering).

While these considerations are important if one wishes to measure an unobscured single-phonon meV-IXS spectrum that one could forward-model (with the intention of extracting phonon information, for example), we should note that if one simply intends to derive a \textit{temperature}, it is not actually necessary to avoid the multiphonon scattering regime. The principle of detailed balance [Eq.~\ref{eq:detailed-balance}] is completely agnostic to the source of the scattering, be it single-phonon, multi-phonon, or some mixture thereof; were the temperature-extraction technique described in Sec.~\ref{sec:laplace} to be applied to a spectrum dominated by, say, four-phonon scattering processes, it would still yield the correct thermodynamic temperature. This is true, at least, provided that the detector has sufficient coverage to collect the full range of allowed photon energy transfers, which is $\pm\ell k_B\Theta_D$ for $\ell$-phonon scattering processes. As quantified by Dornheim \textit{et al.}, if the spectrum is truncated, the expression for the minimum of $F(\mathbf{q},\tau)$ [Eq.~(\ref{eq:laplace-minimum})] becomes inexact, and the inferred temperature can drift significantly from its true value\cite{Dornheim2023}. Given that the spectral window of the diced crystal analyzer (DCA) detectors installed in their standard meV-IXS configuration at EuXFEL is around 460~meV\cite{Wollenweber2021} ($\approx17k_B\Theta_D$ for the ambient Cu studied here), such truncation effects are very unlikely to manifest even somewhat beyond the single-phonon scattering regime for all the but the very hardest elements. The multiphonon regime is therefore viable for temperature measurements, and worthy of further investigation.

The ultimate limit of multiphonon scattering is the single-particle regime. At high enough scattering angles and/or temperatures [i.e., at large $M(\mathbf{q})$], the number of phonons that participate in a typical inelastic scattering event becomes so great that the x-ray photon energy executes something akin to a random walk, resulting in a Gaussian distribution of energy transfers. This same behavior can be explained by considering the Doppler shift of x-rays reradiated by individual atoms whose velocities are distributed according to a Maxwell-Boltzmann distribution. In other words, the collective atomic behavior embodied by the crystal structure and the phonons perturbing it `drop out' of the measured spectrum; the only property on which the shape of the meV-IXS spectrum depends is the temperature. Ultrafast temperature measurements in the single-particle regime have been realised on isochorically heated matter\cite{White2025}, but not, at time of writing, on dynamically compressed material. The price of operating in this non-collective regime is that one cannot measure both temperature \emph{and} phonon properties\cite{McBride2018,Descamps2020a}, as one can in the low-to-intermediate-$q$ regime.

\subsection{Elastic scattering}
The main argument of our study is based around the premise that scattering from the ablator is unavoidable at low $q$. This is true of the commonly used, inexpensive polymer ablators Kapton-B and Parylene-N, whose lack of long-range atomic order means they diffract in all directions (modulated by the polarization factor), leaving no `hiding place' in reciprocal space. Rather than moving to an intermediate $q$-regime to suppress this directionless scattering, as suggested here, a second solution is to use a \emph{crystalline} low-$Z$ ablator (e.g., single-crystal diamond\cite{Rygg2012}) whose diffraction is highly localized in $\mathbf{q}$-space, and therefore avoidable. If the mounting orientation of the ablator were known and consistent across all targets, one could probe at a point in reciprocal space (perhaps even within the first Brillouin zone) remote from the crystalline ablator's reciprocal lattice vectors where its elastic scattering is locally minimal\cite{Descamps2020a}. To estimate for a given target geometry and composition whether the sample's inelastic scattering is strong enough to overcome the wings of the diamond Bragg peaks could be done straightforwardly using the model in Sec.~\ref{sec:theory}. Whether or not producing many hundreds of identically mounted targets each bearing a relatively costly diamond ablator is actually practicable is another question.

The second elastic scattering source that must be avoided is that of the crystalline sample itself. Throughout, we have calculated spectra at discrete $\mathbf{q}$-values. This is to neglect the physical size of the detectors that would in reality collect photons suffering a finite range of momentum transfers. If the window of $\mathbf{q}$-values over which each detector integrates is too wide, it may be impossible to fit between the diffraction peaks, i.e., the detector will inevitably collect intense Bragg scattering that completely overwhelms the inelastic signal. For the current meV-IXS setup at EuXFEL, in which inelastically scattered 7.492~keV x-rays are collected by 10-cm-wide DCA detectors placed 100~cm from the laser-target interaction region,\cite{Wollenweber2021} the integration range is as great as 0.4~\AA\textsuperscript{$-1$}. Inspecting Fig.~\ref{fig:goldilocks}, we see that while the basin between the $(200)$ and $(220)$ peaks is easily wide enough to accommodate such a window, placing the detector so as to probe between the $(220)$ and $(311)$ peaks [separated by only 0.85~\AA\textsuperscript{$-1$}] could be risky, particularly once the Bragg peaks widen during dynamic compression. When operating outside the first Brillouin zone, then, thought must be given to the best choice of inter-Bragg region in which to collect meV-IXS spectra. Perhaps the greatest threat to our strategy is the sample layer being only partially compressed at the x-ray probe time: unwanted Bragg peaks from the remaining \emph{ambient} material at the rear surface of the target may (depending on the shock pressure) land in the habitable zone for the \emph{compressed} material, causing a huge additional source of elastic scattering that could scupper a temperature measurement. It is therefore vital that the target be probed as close as possible to `breakout' (the moment the compression wave reaches the rear surface of the sample layer).

Connected to this same theme is the role of crystallographic defects. Under dynamic compression, plasticity agents such as dislocations\cite{Wehrenberg2017}, stacking faults\cite{Sharma2020} and deformation twins \cite{Turneaure2018,Wehrenberg2017} are created in vast numbers. These defects not only strongly disrupt atomistic order in their immediate vicinity, but they diminish their host crystal's long-range order via the elastic strain fields and localized crystal rotation\cite{Wehrenberg2017,Heighway2021,Heighway2022} they cause. The result is an increase in the elastic scattering cross-section in the wings of and between the Bragg peaks. Our understanding of exactly how dynamic-compression-induced defect networks redistribute elastic scattering intensity around reciprocal space is currently limited. However, a small number of studies that have synthesized structure factors from large-scale, classical molecular dynamics (MD) simulations of shock-compressed metals have already indicated that their enhanced inter-Bragg elastic scattering can easily compete with\cite{Heighway2024} or even overwhelm entirely\cite{Karnbach2021} inelastic scattering, depending on the nature of the defects generated and where in reciprocal space one looks. Systematic investigation of the role of defects in both TDS- and meV-IXS-based temperature measurements of dynamically compressed matter is an essential route for further work, and one for which MD simulations will be absolutely indispensable.

\section{\label{sec:conclusion}Conclusion}
We have put forward a crystallographic-texture--aware model for the momentum- and energy-resolved, single-phonon inelastic x-ray scattering cross-section from a cubic polycrystal. We have used this model to predict the structure of millielectronvolt-scale x-ray scattering spectra from representative dynamic-compression targets comprising a polymer ablator and a polycrystalline sample layer. We have shown that the temperature of the polycrystal can be reliably derived from meV-IXS spectra collected outside the first Brillouin zone where ablator scattering is weak (in spite of their considerably more complex, texture-dependent structure) via the principle of detailed balance. Our results underscore the need to expand the region of reciprocal space over which we consider conducting meV-IXS--based temperature measurements of dynamically compressed matter. The largely uncharted intermediate-$q$ regime (characterized as that in which single-phonon, umklapp IXS dominates) may prove to be the natural home for such measurements, and it is therefore important that we explore both the opportunities and the challenges awaiting us there.

\section*{Supplementary Material}
See the Supplementary Material for a description of the computationally efficient approximation of the shape function for a spherical grain that we used to calculate elastic scattering cross-sections.

\section*{Acknowledgments}
P.\ G.\ H. and J.\ S.\ W.\ gratefully acknowledge the support of the EPSRC under research grant EP/X031624/1. P.\ G.\ H.\ would like to thank D.\ J.\ Peake and T.\ Stevens for the informative discussions that contributed to this study, and further expresses gratitude to E.\ E.\ McBride for their invitation to the DiPOLE 100-X meV-IXS community experiment (Proposal 7902) at EuXFEL, a highly informative experience that catalyzed much of this work.

\section*{Author declarations}
The authors have no conflicts of interest to disclose.

\section*{Data availability}
The synthetic data that support the findings of this study are available from the corresponding author upon reasonable request.

\section*{References}
%

\end{document}


\title{Supplementary Material: Ultrafast temperature measurements of dynamically compressed matter using meV inelastic x-ray scattering beyond the first Brillouin zone}

\author{P.G. Heighway~\orcidlink{0000-0001-6221-0650}}\email{patrick.heighway@physics.ox.ac.uk}
\affiliation{Department of Physics, Clarendon Laboratory, University of Oxford, Parks Road, Oxford OX1 3PU, UK\looseness=-1}

\author{J.S. Wark~\orcidlink{0000-0003-3055-3223}}
\affiliation{Department of Physics, Clarendon Laboratory, University of Oxford, Parks Road, Oxford OX1 3PU, UK\looseness=-1}

\date{\today}

\begin{abstract}
This Supplementary Material describes the computationally efficient approximation of the shape function for a spherical grain that we used to calculate elastic scattering cross-sections in the Main Article.
\end{abstract}

\maketitle

The expression we use for the elastic component of the static structure factor (SSF) of a single, monatomic defect-free, finite-temperature crystallite takes the form
\begin{equation}
    s_0(\mathbf{q}) = \sum_j s_0(\mathbf{q}|\mathbf{G}_j)\ ,
\end{equation}
where the sum runs over all reciprocal lattice vectors $\{\mathbf{G}_j\}$. The individual contribution to the SSF of a single reciprocal lattice vector takes the form
\begin{equation}
    s_0(\mathbf{q}|\mathbf{G}) = N_a f^2(\mathbf{q})e^{-2M(\mathbf{q})}J(\mathbf{q}-\mathbf{G})\ ,
\end{equation}
where $N_a$ is the number of atoms in the crystallite, $f(\mathbf{q})$ is the atomic form factor, $M(\mathbf{q})$ is the Debye-Waller factor, and $J(\mathbf{k})$ is the \textit{shape function}. This final factor describes the delocalization of the SSF around each reciprocal lattice vector owing to the finite size of the crystallite.

To a reasonable approximation, the shape function is proportional to the absolute square of the Fourier transformation of the binary function $B(\mathbf{r})$ delineating the volume occupied by the crystallite. For a spherical grain of radius $R$,
\begin{equation}
    B(\mathbf{r}) = \begin{cases}
        1\quad \text{for}\quad 0\le r\le R\ , \\
        0\quad \text{otherwise}\ .
    \end{cases}
\end{equation}
In this case, the Fourier transform of $B(\mathbf{r})$ can be calculated analytically:
\begin{align}
    \tilde{B}(\mathbf{k}) &= \left | \iiint_{\mathbb{R}^3} d^3\mathbf{r}\,B(\mathbf{r}) e^{-i\mathbf{k}\cdot\mathbf{r}}\right |^2 \\
    &= \left | \int_0^{2\pi}d\phi\int_0^R dr\,r^2\int_0^\pi d\theta\,\sin\theta\,e^{-ikr\cos\theta} \right |^2 \\
    &= \left | \frac{4\pi}{k} \int_0^R dr\,r\sin(kr) \right |^2 \\
    &= \left | 4\pi R^3\left[\frac{\sin(kR) - kR\cos(kR)}{(kR)^3}\right] \right |^2\ .
\end{align}
To ensure the absolute elastic scattering cross-section is predicted correctly, we require that the shape function satisfy the normalization condition
\begin{equation}
    \iiint_{\mathbb{R}^3} d^3\mathbf{k}\,J(\mathbf{k}) = N_b\left(\frac{2\pi}{a}\right)^3\ ,
\end{equation}
where $N_b$ is the number of atoms in the conventional unit cell whose cubic dimension is $a$. For a face-centered cubic (fcc) crystal $(N_b=4)$, the correctly normalized shape function reads
\begin{align}
    \label{seq:shape-function}J(\mathbf{k}) &= \frac{6}{\pi^2}\left(\frac{2\pi R}{a}\right)^2\left[\frac{\sin(kR) - kR\cos(kR)}{(kR)^3}\right]^2 \\
    &= \frac{6}{\pi^2}\left(\frac{2\pi R}{a}\right)^2\hat{J}(kR)\ .
\end{align}

When we examine the structure of the shape function, which is encoded by the function
\begin{equation}\label{seq:Jx}
    \hat{J}(x) = \left(\frac{\sin x - x\cos x}{x^3}\right)^2\ ,
\end{equation}
we see that it vanishes wherever $x=\tan x$. While this oscillating, periodic behavior is of course physically correct for a perfectly spherical grain, when we use Eq.~(\ref{seq:shape-function}) as-is to predict the elastic scattering cross-section of a polycrystal, we produce diffraction patterns that show rapid oscillations between the Bragg peaks. These oscillations are particularly dramatic inside the first Brillouin zone, where the dominant contributions to the structure factor come from each crystallite's $(000)$ reciprocal lattice vector, each of which contributes identically to $s_0(\mathbf{q})$. This artificial fine structure is a severe distraction. Rather than trying to eliminate it by simulating spherical grains with a distribution of sizes, we instead construct a monotonically decreasing approximation of $\hat{J}(x)$ that vanishes nowhere. To do so, we consider the behavior of the dimensionless shape function at small and large $x$, i.e., close to and far from the Bragg peak whose shape it represents.

\begin{figure}
    \centering
    \includegraphics{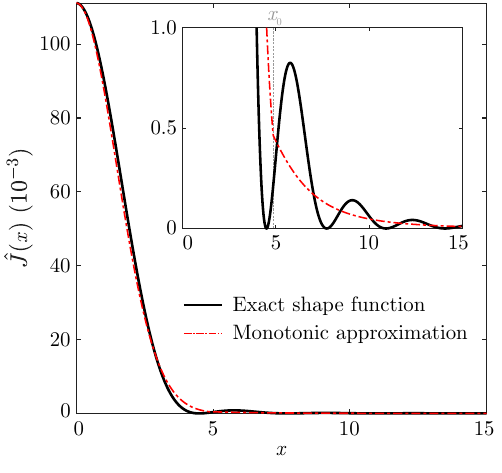}
    \caption{Piecewise approximation [red, Eq.~(\ref{seq:Jxapprox})] of the dimensionless shape function $\hat{J}(x)$ [black, Eq.~(\ref{seq:Jx})]. Inset shows decay behavior in the wing of the peak. A kink is visible at $x_0=4.8548$ where the Gaussian and algebraic intervals of the approximate shape function meet.}
    \label{sfig:shape-function}
\end{figure}

For small $x$ ($\lesssim3$), the shape function strongly resembles a Gaussian. We find that a good approximation to the dimensionless shape function in this domain is
\begin{equation}
    \hat{J}(x)\approx\frac{1}{9}\exp\left(-\frac{x^2}{4.3}\right)\ .
\end{equation}
Conversely, the long-range decay behavior of $\hat{J}(x)$ at large $x$ is easily shown to be
\begin{equation}
    \hat{J}(x)\approx\frac{1}{2x^4}\ .
\end{equation}
The piecewise continuous function that reproduces both the short-range Gaussian behavior and long-range algebraic decay that we choose reads
\begin{equation}\label{seq:Jxapprox}
    \hat{J}_{\text{approx}}(x)=
    \begin{cases}
        \frac{1}{9}\exp\left(-\frac{x^2}{4.3}\right)\quad\text{for}\quad 0\le x\le x_0\ , \\
        \frac{1}{1050+2x^4}\quad\quad\quad\ \text{otherwise}\ ,
    \end{cases}
\end{equation}
where $x_0 = 4.8548$. The exact and approximate dimensionless shape functions are compared in Fig.~\ref{sfig:shape-function}.

Note that this function is not smooth at the point $x=x_0$ where its Gaussian and algebraic components meet, which, in real terms, occurs in the `near-wing' of the Bragg peak. This unphysical behavior is not a problem for our purposes, as millielectronvolt inelastic x-ray scattering (meV-IXS) spectra are only ever collected far from the Bragg peaks that would otherwise completely overwhelm the inelastic scattering of interest.